\documentclass[10pt,prb,aps,twocolumn,showpacs,floatfix,eqsecnum]{revtex4}

\usepackage{graphicx}
\usepackage{amssymb}
\usepackage{subfigure}
\usepackage{color}
\usepackage{amsmath}   
\newcommand{\up}{\uparrow}
\newcommand{\dn}{\downarrow}
\newcommand{\half}{\hbox{$\frac{1}{2}$}}

\newcommand{\MEMO}[1]{{}}

\newcommand{\quarter}{\hbox{$\frac{1}{4}$}}
\newcommand{\beq}{\begin{equation}}
\newcommand{\eeq}{\end{equation}}
\newcommand{\Ftot}{{F_{\rm tot}}}
\newcommand{\hav}{{\bar h}}
\newcommand{\hh}{{\bar h}}
\newcommand{\hraw}{{h}}
\newcommand{\qq}{{\bf q}}
\newcommand{\kk}{{\bf k}}
\newcommand{\QQ}{{\bf Q}}
\newcommand{\rr}{{\bf r}}
\newcommand{\Dxxcrit}{{D_{xx}^{\rm crit}}}

\begin{document}

\title{Properties of Resonating-Valence-Bond Spin Liquids and Critical Dimer Models}

\author{Ying Tang}
\affiliation{Department of Physics, Boston University, 590 Commonwealth Avenue, Boston, Massachusetts 02215, USA}

\author{Anders W. Sandvik}
\affiliation{Department of Physics, Boston University, 590 Commonwealth Avenue, Boston, Massachusetts 02215, USA}

\author{Christopher L. Henley}
\affiliation{Laboratory of Atomic and Solid State Physics, Cornell University, Ithaca, New York 14853-2501, USA}

\begin{abstract}
We use Monte Carlo simulations to study properties of Anderson's resonating-valence-bond (RVB) spin-liquid state on 
the square lattice (i.e., the equal superposition of all pairing of spins into nearest-neighbor singlet pairs) and compare 
with the classical dimer model (CDM). The latter system also corresponds to the ground state of the Rokhsar-Kivelson 
quantum dimer model at its critical point. We find that although spin-spin correlations decay exponentially in the RVB,
four-spin valence-bond-solid correlations are critical, qualitatively like the well-known dimer-dimer correlations 
of the CDM, but decaying more slowly (as $1/r^\alpha$ with $\alpha \approx 1.20$, compared with $\alpha=2$ for the CDM).
We also compute the distribution of monomer (defect) pair separations, which decay by a {\it larger} exponent in the RVB 
than in the CDM. We further study both models in their different winding number sectors and evaluate the relative weights 
of different sectors. Like the CDM, all the observed RVB behaviors can be understood in the framework of a mapping to a 
``height'' model characterized by a gradient-squared stiffness constant $K$. Four independent measurements consistently
show a value $K_{\rm RVB} \approx 1.6 K_{\rm CDM}$, with the same kinds of numerical evaluations of $K_{\rm CDM}$ giving results in 
agreement with the rigorously known value $K_{\rm CDM}=\pi/16$. The background of a nonzero winding number gradient $W/L$ introduces
spatial anisotropies and an increase in the effective $K$, both of which can be understood as a consequence of anharmonic 
terms in the height-model free energy, which are of relevance to the recently proposed scenario of ``Cantor deconfinement'' 
in extended quantum dimer models. In addition to the standard case of short bonds only, we also studied ensembles in which 
fourth-neighbor (bipartite) bonds are allowed, at a density controlled by a tunable fugacity, resulting (as expected) 
in a smooth reduction of $K$.
\end{abstract}

\date{\today}


\maketitle

\section{INTRODUCTION}

The two-dimensional (2D) resonating-valence-bond (RVB) spin-liquid state introduced by Anderson has been studied extensively 
during the past two decades, with the hope that it (when doped) might provide an opportunity to understand high-temperature 
superconductivity in cuprates.\cite{Anderson} Such RVB states, which do not feature any long range magnetic order or 
broken lattice symmetries (but are believed to exhibit non-local, topological order \cite{Read,Bonesteel}) are also of 
broader interest in the context of frustrated magnetism, where they were first considered.\cite{Fazekas} In studies of
specific Hamiltonians, RVB states can be considered as variational ground states. The extreme RVB state built out of only 
the shortest possible (nearest-neighbor) valence bonds (singlets), with equal weights for all bond configurations (which
in the case considered here will be on the square lattice), does not have any adjustable parameters (as long as the signs of 
the wave function are not considered---in the standard RVB all coefficients are equal and positive). One can also parametrically 
introduce longer bonds in amplitude-product states.\cite{Liang} In two dimensions these states are spin liquids if the amplitudes 
decay sufficiently rapidly (exponentially or as a high power) with the bond length. We report here extensive studies
of the RVB state, with only short (length $1$) bonds, as well as in the presence of a fraction of bonds (the second bipartite 
ones of length $\sqrt{5}$).

The search for Hamiltonians with RVB ground states has been an ongoing challenge during the past two decades. One way to approach the problem is through 
quantum dimer models (QDM), in which the internal singlet structure of the 
valence bonds is neglected. The valence bonds are replaced by hard-core 
dimers, and different dimer configurations are considered as orthogonal 
states.\cite{RK1} The effective Hamiltonians in this space, which describe 
the quantum fluctuations of the dimers, can have crystalline dimer order 
[corresponding to a valence-bond-solid (VBS) in the spin system] or be 
disordered (corresponding to a spin liquid). QDMs have many 
interesting and intriguing properties, e.g., the special 
Rokhsar-Kivelson (RK) points at which the wave-function of a dimer model corresponds exactly to the statistical mechanics of 
classical dimers.\cite{RK1,Henley,Henley-RK,Castelnovo} On the square lattice the classical dimer model (CDM) has critical dimer-dimer 
correlations, decaying with distance $r$ as $1/r^2$ (a rigorous result \cite{Stephenson}) which then is also the case at the RK point 
separating two different VBS states on the square lattice. On the triangular lattice, this isolated spin-liquid point with
critical dimer correlations is replaced by an extended liquid phase with exponentially decaying dimer correlations.\cite{Moessner1} 
The same physics can be achieved on the square lattice by introducing dimers between next-nearest-neighbor sites.\cite{Sandvik-Moessner}
We will here also provide some further results for the CDM, in order to elucidate in more detail the relationship between
the RVB and the CDM.

Formally, the QDMs can be related exactly to generalized SU($N$) symmetric spin models.\cite{Read2} In the 
limit of $N\to \infty$ the valence-bond states become exactly orthogonal dimer states. Whether or not the physics of the quantum
dimer models can be extended down to the physically most interesting case of SU($2$) spins is in general not clear (unless
the $N=2$ features are built in from the start, as can be done in {\it generalized QDMs} \cite{Schwandt}). 
Moessner and Sondhi have devised a procedure to mimic a system of large-$N$ spins by decorating an original lattice of $S=1/2$ SU($2$) 
spins with additional spins, and this way a Hamiltonian with spin-liquid ground state can be constructed.\cite{Raman} 
Very recently, Cano and Fendley constructed a Hamiltonian the ground state of 
which is exactly the short-bond RVB state on the 
square lattice (without decoration).\cite{Cano} While this Hamiltonian is a complicated one with multi-spin interactions that 
are unlikely present in real systems, the achievement is important as it shows that local SU($2$) spin models with RVB states 
do in principle exist also on simple lattices.

\subsection{Correlations in RVB and dimer states}

Perhaps surprisingly, very few physical properties of RVB spin liquids have actually been computed. While Monte Carlo
simulations of amplitude-product states on the 2D square lattice were carried out some time ago, only the simple spin-spin 
correlations were calculated.\cite{Liang} They decay exponentially in the case of the short-bond state. On the other hand,
the fact that the dimer-dimer correlations of the CDM (or, equivalently, the 
QDM at the RK point) decay with a
power-law clearly suggests that there should be similar critical correlations also in the RVB state  (if the QDM 
is qualitatively faithful to it). The dimer-dimer correlations of the RVB state are not physical correlations, however, as the 
dimer basis is non-orthogonal and overcomplete. 

In this paper, we use an improved Monte Carlo sampling scheme for valence bonds 
\cite{Sandvik2}  to compute the physical correlation function most closely related to the dimer-dimer correlations of the CDM, 
namely, the four-spin correlation function
\begin{equation}
D_{xx}({\bf r}_{ij}) = \langle B_x({\bf r}_i) B_x({\bf r}_j)\rangle ,
\label{dimer}
\end{equation}
where $B_x({\bf r}_i)$ is a scalar operator defined on a bond,
\begin{equation}
B_x({\bf r}_i) = {\bf S}({\bf r}_i) \cdot {\bf S}({\bf r}_i+\hat {\bf x}),
\label{bond}
\end{equation}
and $D_{yy}$ and $D_{xy}$ can be defined analogously. Here the lattice coordinate of spin $i$ is denoted ${\bf r}_i$ and
$\hat {\bf x}$ is the lattice vector in the x-direction. The operator $B_x({\bf r}_i)$ provides a measure of the singlet 
probability on the bond between site $i$ and its ``right'' neighbor, which is larger on a valence bond (in which case
the operator is diagonal) than between two valence bonds (where the operator is off-diagonal and leads to a rearrangement
of the two valence bonds). It is therefore appropriate to consider $B({\bf r}_i)$ as the ``quantum dimer'' operator
to be used in place of the dimer density $n_x({\bf r}_i) \in \{0,1\}$ in the CDM. Because of the non-orthogonality of the 
valence-bond basis, $D_{xx}({\bf r})$ is not, however, identical to the classical dimer-dimer correlation function. The
two systems and their dimer correlation functions become identical in SU($N$) symmetric generalizations of the RVB when
$N \to \infty$.\cite{Read2} 

We will here show that $D_{xx}(r)$ for the standard $S=1/2$ SU($2$) spins decays much slower 
than the classical correlator, as $1/r^{\alpha}$ with $\alpha \approx 1.20$. These correlations, which are peaked at momenta 
${\bf q}=(\pi,0)$ and ${\bf q}=(0,\pi)$, correspond to critical fluctuations of a columnar valence-bond-solid (VBS). 
The exponent $\alpha < 2$ in the RVB spin liquid corresponds to power-law divergent Bragg peaks, while in the CDM these peaks 
are only logarithmically divergent. As a consequence of the non-orthogonality of the valence-bond basis, the  RVB is, thus, 
significantly closer to an ordered VBS state than is the CDM (or QDM). This result was first reported by
us in a conference abstract \cite{yingabstract} and in an unpublished earlier version of this paper \cite{vers1}, 
and was also found in independent parallel work by Albuquerque and Alet.\cite{albu10} Here we provide further details on 
the dimer correlations and their significance.

We also study systems doped with two monomers and compute the distribution function of the monomer separation. A well
known result for the CDM is that the monomers are deconfined, with the distribution function $M(r)$ decaying with the
separation $r$ as $A(L)/r^{\beta}$, where $\beta=1/2$ and the prefactor $A(L)$ decays with the system size $L$ in such a way 
that the distribution is normalized for all $L$. For the RVB state, we find a more rapid power-law decay, with 
$\beta \approx 0.83$, which still corresponds to deconfined monomers.

It is known that the dimer correlations of the CDM decay as $1/r^2$ also in the presence of longer bipartite bonds (while non-bipartite
bonds leads to a non-critical phase, with exponentially decaying correlations). As we will explain further below and in Appendix B, 
the exponent $\alpha$ in this case does not correspond to these leading correlations, however, but a subleading contribution decaying 
as $1/r^\alpha$ with $\alpha>2$. This exponent and the monomer exponent $\beta$ are non-universal, depending on details of the model 
(the fugacities corresponding to the longer bonds).\cite{Sandvik-Moessner} 
We also study here the RVB including longer bonds (the second 
bipartite bond, which connects fourth-nearest neighbors as considered previously in the CDM \cite{Sandvik-Moessner}) and find that also 
in this case $\alpha$ and $\beta$ change with the concentration of longer bonds. In contrast to the CDM, the leading dimer correlations 
are always (at least for the range of parameters studied here) controlled by $\alpha$, however, since $\alpha<2$ for the RVB.

\subsection{Height representation and topological sectors}
\label{sec:height-intro}

A key notion for relating the various results
on the CDM and (we believe) the RVB model also, 
is that of ``height model,'' or equivalently
a $U(1)$ classical field theory.  This means
that all the long-wavelength behaviors of
the system are captured by a coarse-grained
scalar field $\hh(\rr)$.  The dimer density
operators and monomer defects can all be
expressed in terms of $\hh(\rr)$, and the
weighting of its configurations is proportional
to $\exp(-\Ftot)$, where 
\beq
    \Ftot = \int d^2 \rr \frac{1}{2} K |\nabla \hh(\rr)|^2.
    \label{eq:Ftot-h-main}
\eeq
The height mapping for square-lattice dimers was introduced 
over twenty years ago.\cite{Zheng-Sachdev,levitov,ioffe-larkin}
The use of such a mapping to explain correlation functions
originated earlier (effectively for dimers on a honeycomb lattice)
with Bl\"ote, Hilhorst, and Nienhuis.\cite{Blote}

The key parameter in Eq.~(\ref{eq:Ftot-h-main}) is the
dimensionless stiffness constant $K$.
It can be shown that
the exponents $\alpha$ and $\beta$ measured in our
simulations, as well as 
the coefficients of a ``pinch-point'' singularity
in the dimer-density structure factor, and also
the ratios of the probabilities of different topological
(winding number) sectors, are all functions purely of $K$. The details of the 
height-model construction underlying this result are given in 
Appendix~\ref{app:height}. It will be shown in Sec.~\ref{sec:height-anal}  
that all our measurements based on Monte Carlo simulations of the CDM and RVB 
consistently give the {\it same} value of $K$ for a given model, demonstrating the
validity of the height model. That is expected for the CDM, for which the height approach 
is well known; here we show that it is pertinent to the RVB as well.

\begin{figure}
\centerline{\includegraphics[angle=0,width=7.5cm, clip]{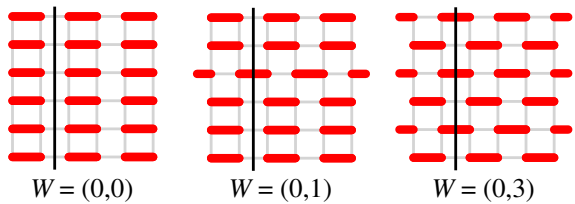}}
\caption{(Color online) Configurations in different winding number sectors, $W=(W_x,W_y)$. Here $W_y$ is given by 
the number of bonds crossing the line drawn in the $y$-direction (since those bonds are at even $y$---shifting the bond configuration
by one step in the $y$-direction leads to $W_y \to - W_y$). The last case is the unique configuration in its winding number sector and 
constitutes the staggered state of the QDM.\cite{RK1}} 
\label{sectors}
\end{figure}

\begin{figure}
\centerline{\includegraphics[angle=0,width=8.25cm, clip]{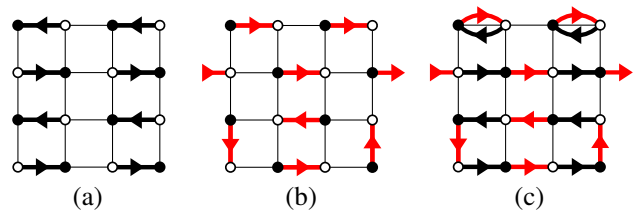}}
\caption{(Color online) (a) Reference state used here for defining the winding number. The direction of the dimers is from sublattice B 
(open circle) to sublattice A (solid circle). (b) An arbitrary valence bond state, with dimers drawn in the opposite direction, from 
sublattice A to sublattice B. (c) The transition graph formed by the reference states in (a) and the arbitrary state in (b). The winding 
numbers correspond to the net fluxes (in units of the system length $L$) defined by traversing the loops formed along the arrows; 
here $\Phi_x=1$ and $\Phi_y=0$, or $\Phi=(1,0)$, which corresponds to winding number $W=(0,1)$ in the definition of Fig.~\ref{sectors}}
\label{winding}
\end{figure}

A related aspect of RVB states and the CDM is that their bond configurations on periodic lattices can be classified according to a 
topological winding number.\cite{RK1} We here define the winding number $W=(W_x,W_y)$ as used in Ref.~\onlinecite{Fradkin2}. Drawing 
a path in the $y$ direction, $W_y$ is the number of $x$-dimers crossed at even $y$ minus the number of such dimer crossing at odd $y$ (see Fig.~\ref{sectors}). 
An equivalent definition~\cite{RK1} uses one of the $W=0$ single-domain states, such as the one in Fig.~\ref{winding}(a), as a 
reference state. As shown in Fig. \ref{winding}(c), a direction can be assigned to loops of the transition graph so that each carries a ``lattice flux''; if we 
call the net fluxes $(\Phi_x,\Phi_y)$, then ($W_x,W_y)=(\Phi_y, \Phi_x)$ [or, depending on exactly which reference state is used and how 
the $y$ coordinates are assigned, we could have ($W_x,W_y)=(\Phi_y, -\Phi_x)$---the signs are normally not important]. This definition can 
be directly extended to systems with long dimers, by associating that flux (which can have both $x$ and $y$ components, for cases where
there are bonds not along the $x$ or $y$ axis) with a line connecting their endpoints. A third definition of the same winding number is 
(proportional to) the net height difference added up along a path crossing the system in the $x$ or $y$ direction, using the rules detailed 
in Appendix ~\ref{app:height}. The possible winding values for an $L\times L$ lattice are  $W_x,W_y \in \{-L/2, -L/2+1, \ldots, L/2\}$.
The equal-weighted (CDM) ensemble is dominated by the winding number sector ${\bf W}=(W_x,W_y)=(0,0)$ [as follows from $\nabla \hh=0$ 
being the minimum of Eq.~(\ref{eq:Ftot-h-main})].

Recently, extended QDMs have been considered, with interaction
terms that can drive the system into ground states with non-zero 
$\nabla h$ in a sequence of commensurate locking transitions.\cite{Fradkin1,Fradkin2} 
Quantum phase transitions involving these states are unusual, exhibiting aspects of deconfinement on a fractal curve of critical points (forming
a Cantor set, which prompted the term ``Cantor deconfinement'' for this class of unconventional transitions). 
This motivates us to also study the CDM and RVB states in different winding number sectors,
which (it turns out) also happens to be an effective probe of the states' topological natures.
In the case of the RVB, states defined within sectors of different winding numbers are not orthogonal, but become orthogonal 
in the limit of the infinite lattice (which we will here demonstrate explicitly based on simulations). 

\subsection{Outline of the paper}

The outline of the rest of the paper is as follows: In Sec.~\ref{basis} we review the essential features of the valence bond basis 
that we use for the RVB-state calculations, in particular how to extract spin correlations. The four-spin correlations are re-derived in 
detail in Appendix~\ref{app:4spin-corr}, in an alternative way to a previous treatment of more general multi-spin interactions.\cite{Beach} 
In Sec.~\ref{algorithm} we discuss Monte Carlo two-bond reconfiguration \cite{Liang} and loop-cluster algorithms for sampling the 
CDM and RVB states. We also discuss the winding numbers and issues related to sampling them either grand-canonically (where there are some 
ergodicity issues in the case of the RVB) or canonically. In Sec.~\ref{results} we present results for the standard case of only 
length-$1$ dimers and valence bonds, as well as extended models with bonds of length $\sqrt5$. In Sec.~\ref{sec:height-anal} the results are 
interpreted in terms of a height model. Detailed derivations of height model predictions are left to Appendix ~\ref{app:height}. 
In Sec.~\ref{vbshisto} we further characterize the nature of the critical VBS fluctuations in terms of the joint probability
distribution of the order parameters for horizontal and vertical bond ordering. We conclude in Sec.~\ref{conclusion} with a brief 
summary and discussion. 

\section{The Valence bond basis}
\label{basis}
We work in the standard bipartite valence bond basis, where a state of $N$ (an even number of) spins on a bipartite lattice,
\begin{equation}
|V_\alpha\rangle = \frac{1}{2^{N/4}}\prod_{i=1}^{N/2} (|\up_i\dn_{\alpha(i)}\rangle-|\dn_i\up_{\alpha(i)}\rangle),
\label{valpha}
\end{equation}
is a product of singlets, where the first spin $i$ of each singlet is on sublattice $A$ and the second spin $\alpha(i)$ is on sublattice $B$.
With the $B$ sites also labeled as $1,\ldots,N/2$, the set $\alpha(1),\ldots,\alpha(N/2)$ is a permutation of these numbers and the label 
$\alpha=1,\ldots (N/2)!$ in $|V_\alpha\rangle$ simply refers to all these permutations. The signs of the expansion coefficients of this state in 
the standard $\up,\dn$ spin basis correspond to Marshall's sign rule for the ground state $|\Psi_0\rangle$ of a bipartite system,\cite{Marshall} 
i.e., 
\begin{equation}
{\rm sign}[\Psi_0(S^z_1,\ldots,S^z_N)]=(-1)^{n_{A\dn}},
\end{equation}
where $n_{A\dn}$ is the number of $\dn$ spins on sublattice A.

\begin{figure}
\center{\includegraphics[width=8.2cm]{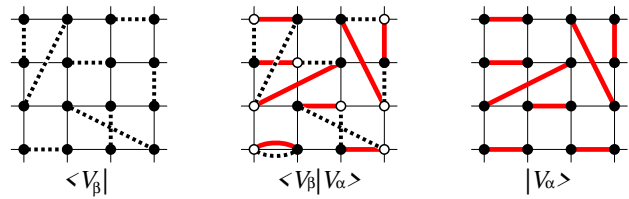}}
\caption{(Color online) Two valence-bond states (left and right) in two dimensions and their transition graph formed by superimposing the two bond 
configurations (center). One of the spin configurations compatible with the transition graph is also shown, with open and solid circles 
for $\up$ and $\dn$ spins. Each loop has two such allowed staggered spin configuration, and the overlap of two valence-bond states is thus 
$\langle V_\beta|V_\alpha\rangle=2^{n_{\alpha\beta}-N/2}$, here with the number of loops $n_{\alpha\beta}=4$ and the number of spins $N=16$.}
\label{olap}
\end{figure}

An {\it amplitude-product} state is a superposition of valence bond states,
\begin{equation}
|\Psi\rangle = \sum_\alpha \psi_\alpha |V_\alpha\rangle,
\end{equation}
where the expansion coefficients are products of amplitudes $h({\bf r}_{\alpha,i})$ corresponding to the ``shape'' of the bonds
(the bond lengths in the $x$ and $y$ direction in the case of a 2D system);
\begin{equation}
\psi_\alpha =\prod_{i=1}^{N/2} h({\bf r}_{\alpha,i}).
\label{hprod}
\end{equation}
Our main focus here will be on the extreme RVB state made up of only bonds of length 1 (one lattice constant), in which case the expansion coefficients 
$\psi_\alpha$ are the same for all configurations. We will also later study states including the bipartite bonds of length $\sqrt{5}$ lattice constants, 
examples of which are seen in Fig.~\ref{olap}. 
The discussion here and in Sec.~\ref{algorithm} will be framed around 
generic bipartite amplitude-product states, with no restriction on the bond lengths.

\subsection{Transition graphs}

An important concept in the valence bond basis is the transition graph formed when the bond configurations of the two states are 
superimposed.\cite{Sutherland,Liang} This is illustrated in Fig.~\ref{olap}. The overlap $\langle V_\beta|V_\alpha\rangle$ between two valence-bond 
basis states can be simply expressed in terms of the number $n_{\alpha\beta}$ of loops in the transition graph. 

The easiest way to calculate the 
overlap is to go back to the standard basis of $\up$ and $\dn$ spins, so that
\begin{eqnarray}
&&\langle V_\beta|V_\alpha\rangle = \frac{1}{2^{N/2}}
\sum_{S^z_\alpha}\sum_{S^z_\beta}
 (-1)^{n_{\alpha,A\dn}+n_{\beta,A\dn}} \times \label{overlapzz} \\
&&~~~~~~~~~~~~~~
\bigl \langle S^z_{\beta 1},\ldots,S^z_{\beta N}|S^z_{\alpha 1},\ldots,S^z_{\alpha N} \bigr\rangle, \nonumber
\end{eqnarray}
where $S^z_\alpha$ and $S^z_\beta$ denote spin configurations compatible with the bond configurations $V_\alpha$ and $V_\beta$, i.e., those that have spins
$\up\dn$ or $\dn\up$ on each bond. Terms with any occurrence of $S^z_{\alpha i}\not= S^z_{\beta i}$ of course vanish, and the double sum, thus, simply 
counts the number of spin configurations common to the two bond configurations. Since the spins on each bond are antiparallel, the spins along a loop 
of alternating $V_\alpha$ and $V_\beta$ bonds (i.e., the loops in the transition graph) must alternate in a staggered, $\up\dn\up\dn\ldots$, pattern. 
There are two such configurations for each loop. The total number of contributing spin configurations is therefore $2^{n_{\alpha\beta}}$, giving the overlap 
\begin{equation}
\langle V_\beta|V_\alpha\rangle = 2^{(n_{\alpha\beta} - N/2)},
\label{overlap}
\end{equation}
which replaces the orthonormality condition $\langle \beta |\alpha\rangle = \delta_{\alpha\beta}$ for an orthonormal basis. For bond tilings 
$V_\alpha=V_\beta$, we have $n_{\alpha\beta}=N/2$ and the overlap equals unity.

In calculations with superpositions  $|\psi\rangle$ of valence-bond states, such as amplitude-product states, it is often not practical to 
normalize the states. It is convenient to write operator expectation values in the form
\begin{eqnarray}
\langle \Psi |O|\Psi\rangle & = & 
\frac{\sum_{\alpha\beta} \psi_\beta \psi_\alpha \langle V_\beta |O|V_\alpha\rangle}
{\sum_{\alpha\beta} \psi_\beta \psi_\alpha \langle V_\beta |V_\alpha\rangle} \nonumber \\
&= & \frac{\sum_{\alpha\beta} \psi_\beta \psi_\alpha \langle V_\beta |V_\alpha\rangle 
\frac{\langle V_\beta |O|V_\alpha\rangle}{\langle V_\beta |V_\alpha\rangle}}
{\sum_{\alpha\beta} \psi_\beta \psi_\alpha \langle V_\beta |V_\alpha\rangle}.
\label{psivbpexpvalue}
\end{eqnarray}
Defining the weight $W_{\alpha\beta}$ for the combined bond configuration $V_\alpha,V_\beta$ and the normalized
matrix element $O_{\alpha\beta}$ according to
\begin{eqnarray}
W_{\alpha\beta} & = & \psi_\beta \psi_\alpha \langle V_\beta |V_\alpha\rangle, \label{psivbweight} \\
O_{\alpha\beta} & = & \frac{\langle V_\beta |O|V_\alpha\rangle}{\langle V_\beta |V_\alpha\rangle}, \label{psivbestimator}
\end{eqnarray}
the expectation value takes the form appropriate for use with the Monte Carlo sampling methods that we will discuss below in Sec.~\ref{algorithm};
\begin{equation}
\langle \Psi |O|\Psi\rangle = 
\frac{\sum_{\alpha\beta} W_{\alpha\beta} O_{\alpha\beta}}{\sum_{\alpha\beta} W_{\alpha\beta}}.
\label{psivbpexpvaluemc}
\end{equation} 
The weight $W_{\alpha\beta}$, which is used in sampling the states in Monte Carlo simulations, is positive-definite when we consider wave 
functions satisfying Marshall's sign rule, i.e., the amplitudes $h({\bf r}_{\alpha,i}) \ge 0$ in Eq.~(\ref{hprod}). 

Like the overlap of the valence-bond states, the matrix elements of operators of interest can typically also be expressed in terms of the loops 
of the transition graph of the bond configuration $V_\alpha,V_\beta$. We discuss spin and dimer correlations next. 

\subsection{Correlation functions}

The standard spin-spin correlation function is most easily obtained by reintroducing the spins in the transition graph, as illustrated in 
Fig.~\ref{olap}. We can then use the fact that 
\begin{equation}
\langle V_\beta| {\bf S}_i \cdot {\bf S}_j |V_\alpha \rangle = 3 \langle V_\beta |S^z_i S^z_j|V_\alpha \rangle,
\label{szz}
\end{equation}
where the latter is diagonal and easy to compute in the $z$-spin basis. When summing over the allowed spin states, i.e., the two ``orientations'' of 
each loop (for a total of $2^{n_{\alpha\beta}}$ spin states), it is clear that $S^z_i S^z_j$ averages to zero if $i$ and $j$ are in different loops, 
whereas for $i,j$ in the same loop we get $\pm \frac{1}{4}\langle V_\beta|V_\alpha\rangle$, with the sign depending on whether the spins are in the 
same ($+$ sign) or different ($-$ sign) sublattices. Introducing the notion $(i,j)_L$ for two spins in the same loop and $(i)_L(j)_L$ for spins
in different loops, we can write the matrix element ratio in Eq.~(\ref{psivbpexpvaluemc}) corresponding to the spin correlation function as
\begin{equation}
\frac{\langle V_\beta |{\bf S}_i \cdot {\bf S}_j|V_\alpha\rangle}{\langle V_\beta |V_\alpha\rangle} =
\left \lbrace \hskip-1mm
\begin{array}{ll}
                   0, & ~~~(i)_L(j)_L  \\
\frac{3}{4}\phi_{ij},  &  ~~~(i,j)_L,
\end{array}\right.
\label{scorr2}
\end{equation}
where $\phi_{ij}$ is the staggered phase factor;
\begin{equation}
\phi_{ij} = \left\lbrace \hskip-1mm\begin{array}{l}
-1,~~~\hbox{for $i,j$ on different sublattices}, \\
+1,~~~\hbox{for $i,j$ on the same sublattice}.
\end{array}\right.
\end{equation}
While the loop-expression Eq.~(\ref{scorr2}) for the simple spin-spin correlation function is well known,\cite{Liang, Sutherland} the general form of a four-spin 
correlation (of which the dimer-dimer correlator of interest here is a special case) was only derived recently.\cite{Beach} 
In Appendix~\ref{app:4spin-corr} we discuss this derivation in a slightly different way, which is less convenient when generalizing to higher-order 
correlators (which was also done in Ref.~\onlinecite{Beach}), but more transparent in the case of the four-spin correlator. The resulting general 
formula for any non-zero four-spin matrix element is 
\begin{eqnarray}
&&\frac{\langle V_\beta |({\bf S}_k \cdot {\bf S}_l)({\bf S}_i \cdot {\bf S}_j) |V_\alpha\rangle} {\langle V_\beta |V_\alpha\rangle}  = \nonumber \\
&&~~\left\lbrace \hskip-3mm
~~~\begin{array}{ll}
(\frac{9}{16}-\frac{3}{4}\delta^{ij}_{kl})\phi_{ij}\phi_{kl},& ~~~(i,j,k,l)_L,  \\
\frac{9}{16}\phi_{ij}\phi_{kl}, & ~~~(i,j)_L(k,l)_L, \\
\frac{3}{16}\phi_{ij}\phi_{kl}, & ~~~(i,k)_L(j,l)_L, \\
\frac{3}{16}\phi_{ij}\phi_{kl}, &~~~(i,l)_L(j,k)_L.
\end{array}\right.~~~~~~~~~
\label{scorr4}
\end{eqnarray}
Here we have generalized the notation of Eq.~(\ref{scorr2}) for how the sites are distributed among loops in a straight-forward way, with indices within the same 
parentheses belonging to the same loop. In the case of the single-loop contribution, $(i,j,k,l)_L$, the term $\delta^{ij}_{kl} \in \{0,1\}$ depends on the order 
of the four indices within the single loop, as specified in Eq.~(\ref{deltaijkl}) of Appendix~\ref{app:4spin-corr}. 

\section{Monte Carlo Algorithms}
\label{algorithm}

A simple but powerful Monte Carlo sampling algorithm for amplitude-product states based on reconfiguration of bond pairs was presented some times ago by 
Liang {\it et al.},\cite{Liang} who used this method to study the spin-spin correlations in amplitude-product states with several different forms of the 
amplitudes (exponentially or power-law decaying with the length of the bond). A more efficient algorithm using loop updates was developed recently which 
operates in a combined basis of both valence bonds and spins.\cite{Sandvik2} The two-bond update, as well, can be made more efficient by working in this 
combined basis. Here we briefly review these two algorithms, and 
also discuss the topological winding numbers that can be used to 
classify the bond configurations.

\subsection{Combined bond-spin basis}

Monte Carlo sampling of valence bonds involves making some change in the bra and ket bond configurations $V_\alpha$ and $V_\beta$, and accepting or rejecting
the update based on the change in the sampling weight Eq.~(\ref{psivbweight}), according to some scheme satisfying detailed balance. Working with the standard 
non-orthogonal valence bond basis and using the Metropolis algorithm, we need to compute the weight ratio appearing in the acceptance probability
\begin{equation}
P_{\rm accept}={\rm min}\left [\frac{W_{\alpha^\prime\beta^\prime}}{W_{\alpha\beta}},1\right ],
\label{pacc}
\end{equation}
where the primes indicate the new states after some changes have been made in either bond configuration $V_\alpha$ or $V_\beta$ (or both, but typically 
one would change only one state at a time). 

The weight ratio using Eq.~(\ref{psivbweight}) is
\begin{equation}
\frac{W_{\alpha^\prime\beta^\prime}}{W_{\alpha\beta}}=\frac{\psi_{\alpha^\prime}\psi_{\beta^\prime}}{\psi_{\alpha}\psi_{\beta}}
2^{(n_{\alpha^\prime\beta^\prime}-n_{\alpha\beta})}.
\label{wratio}
\end{equation}
For an amplitude-product state, the ratio of the wave function coefficients is trivial, but computing the change 
$n_{\alpha^\prime\beta^\prime}-n_{\alpha\beta}$ in the number of loops in the transition graph can be time consuming, as it 
involves tracing loops that can be long. 

The loops are typically long, $\mathcal{O}(N)$, if there is antiferromagnetic 
long-range order.\cite{Sandvik2} That is not the case for the short-bond RVB states studied in this paper, but nevertheless it 
is more efficient to avoid the loop-counting step. That can simply be done by expressing each singlet in the standard basis of 
$\up$ and $\dn$ spins, and sampling these spin configurations in addition to the bond configurations [and since the spin basis 
is orthonormal, the sampled (non-zero weight) spin configurations must be the same in the bra and the ket]. That is, the 
configurations being sampled consist of a direct product of {\it two} valence bond patterns $V_\alpha$ and $V_\beta$, as well 
as one spin configuration $Z_{\alpha\beta}$ compatible with both $\alpha$ and $\beta$ (i.e. one $\up$ and one $\dn$ spin on each 
bond). Each loop in the transition graph must consist of an alternating string $\up\dn\up...\dn$ and, for every loop, there 
are two choices for this string. Thus, the ratio of the number of spin configurations is equal to the factor 
$2^{(n_{\alpha^\prime\beta^\prime}-n_{\alpha\beta})}$ in Eq.~(\ref{wratio}). The Monte Carlo sampling of the spin configurations compatible 
with the bond configurations therefore automatically takes care of the factor $2^{n_{\alpha\beta}}$ in Eq.~(\ref{wratio}), with no need to generate a transition 
graph or count loops. For more details of the arguments leading to this 
conclusion, see Ref.~\onlinecite{Sandvik2}. 

\subsection{Monte Carlo sampling}

Here we outline the two different bond sampling algorithms that we used, each of which comes in a simple version for the CDM, 
as well as a generalization for the combined spin-bond basis for the RVB amplitude-product states. In the case of the RVB, the
spin configurations also have to be updated. We also introduce a simple extension to sample states with monomers (empty sites). 

\begin{figure}
\centerline{\includegraphics[angle=0,width=7.5cm, clip]{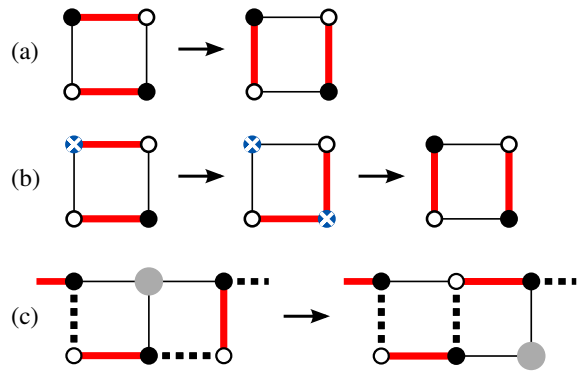}}
\caption{(Color online) Monte Carlo updates for the RVB state in the combined spin-bond basis. Open and solid circles represent 
$\up$ and $\dn$ spins. In the basic moves (a) and (b), only one of the two two valence bond configurations is affected at a time.
(a) A simple two-bond update. Choosing two sites on the same sublattice, the two bonds connected to them can be reconfigured in a 
unique way. If the spins are compatible with the $\up,\dn$ singlet restriction, this update can be accepted. (b) Loop-cluster update. 
Choosing an arbitrary starting site (in this example in the left-upper corner) two defects (a site with no dimer or two dimers 
connected to it, both indicated with an $\times$) are generated by moving the end of the dimer on the initial site to another site 
which satisfies the bond-length constraint (here, in the extreme short-bond RVB, the length is always one) and the spin-singlet 
compatibility (anti-parallel spins on the bond). The dimer that was previously connected to this site is then moved away from the 
double-bond defect to another site. This process continues until a bond returns back to ``annihilate'' the original empty-site defect, 
which here happens already after two bond moves [the last step in (b)]. In both (a) and (b), we only show the bonds of the configuration 
involved in this update. (c) Monomer update. Monomers are shown as larger circles and must appear in the same locations in the state 
$|V_\alpha\rangle$ and $|V_\beta\rangle$, the bonds of both of which are shown here (as solid and dashed lines). In addition to the 
two-bond or loop update of the bonds, monomers can move to a site on the same sublattice by also moving a bond which is common 
to the two valence bond states.}
\label{update}
\end{figure}

The two updating algorithms are summarized using simple examples with short bonds in Figs.~\ref{update}(a,b), with (c) showing 
the extension needed for also sampling monomer configurations. For either algorithm, updates are alternated between the 
ket and bra configurations, and there is an additional step for updating the spin configuration, where all the spins belonging
to randomly chosen individual loops in the transition graph are flipped.

\subsubsection{Two-bond update}

For the two-bond update, as in Ref.~\onlinecite{Liang} we choose two sites on the same sublattice (normally a next-nearest-neighbor 
site pair) and exchange their dimers in the unique way maintaining the $A-B$ sublattice connectivity, as shown in Fig.~\ref{update}(a). The 
update can be accepted only if the spin configuration is compatible with the new bond structure, i.e., only antiparallel spins are connected 
by the bonds. In the case of the extreme short-bond RVB, an allowed new configuration is always accepted, as the wave function ratio 
in Eq.~(\ref{wratio}) trivially equals one, whereas in general, when longer bonds are present, a ratio involving the amplitudes of two bonds 
has to be computed to determine the Metropolis acceptance rate Eq.~(\ref{pacc}). 

The algorithm for the CDM is simpler, as there is no spin state in that case. In the case of short bonds, an update of two bonds [flipping a pair of 
parallel bonds as in Fig.~\ref{update}(a)] is then always accepted, whereas in the presence of longer bonds the acceptance probability involves the ratio
of bond fugacities. We here consider only two bond lengths (nearest neighbor and fourth-nearest neighbor bonds, as shown in Fig.~\ref{olap}), with fugacities $Z_1(i)=1$ and $Z_2(i)$, respectively, for bonds connected to site $i$ (taken to be the sublattice A site, for definiteness). The partition function is then given by 
\begin{equation}
\mathcal{Z_{CDM}}=\sum_{C}Z_2^{n_2(C)}
\end{equation}
where $n_2(C)$ is the number of long bonds in configuration C. The acceptance probability for an update of bonds on sites $i$ and $j$ is
\begin{equation}
P_{\rm acc}={\rm min}\left [\frac{Z_{\rm new}(i)Z_{\rm new}(j)}{Z_{\rm old}(i)Z_{\rm old}(j)},1\right],
\end{equation}
where "old" and "new" correspond to the length-index $1$ or $2$ before and after the bond reconfiguration.

For both the RVB and CDM, this algorithm keeps the system in a sector of fixed winding number, which we can take advantage of if we want 
to study properties in the individual sectors. Suitable starting configurations for different winding number sectors are shown 
in Fig.~\ref{sectors}. 

\subsubsection{Loop update}

If we want the system to wander among the different topological sectors, we instead use the loop-cluster update, which is a simple extension of
a loop update for the CDM.\cite{Adams,Sandvik-Moessner} It is also in general more efficient (exhibits shorter autocorrelation times) than the two-bond 
update for large size system. To start the loop update, we pick a site at random; in the example in Fig.~\ref{update}(b) the top left site. We 
move the dimer connected to it, thus creating two defects in the system. We keep the starting site as a vacancy and move the original dimer of the 
now doubly occupied site to a new site, with certain probabilities satisfying detailed balance, and constrained by the spin configuration so that 
spins are opposite on every dimer. In the case of short bonds only, the probabilities 
are equal for the three new neighbor sites. For the general case where longer bonds are included, we refer to Ref.~\onlinecite{Sandvik-Moessner} for
efficient choices of the probabilities. This update moves the doubly-occupied defect to a new site, which in Fig.~\ref{update}(b) is the lower-right 
site. We keep moving this defect using the above procedures, until it happens that the two defects annihilate each other, which means that bonds have
been moved on a closed loop of sites. A sweep of bond updates is defined as the construction of a fixed number of loops (determined during the
equilibration part of the simulation) which on average result in $\approx N$ moved bonds in both the ket and the bra state.

\subsubsection{Spin update}

After updating the bond configurations with one of the above algorithms, we update the spin configuration by flipping the spins of randomly 
selected loops of the  transition graph (such as those in the middle graph of Figs.~\ref{olap}), with probability $1/2$ for each loop. All the 
loops have to be traversed, by moving between spins according to the bonds (which are stored in the computer as bidirectional links), alternating 
between bonds in the bra and ket state. Each site visited is flagged and no new loops are started from already visited sites. The computational 
cost of a full sweep of such  updates (visiting each site once) is $\mathcal{O}(N)$. 

\subsubsection{Monte Carlo sweep}

A sequence of bond updates in which $\mathcal{O}(N)$ bonds are affected followed by a complete spin update constitutes one Monte Carlo sweep, which 
has a total computational cost $\mathcal{O}(N)$. Note that the sampling algorithm without the spins potentially costs up to $N^2$ steps per sweep, 
since each two-bond update requires loop-traversals to check whether two loops are joined or a single loop is split,\cite{Liang} and the 
loop length can then be up to $\mathcal{O}(N)$ (in a N\'eel state). The same issue pertains to loop updates in the pure valence-bond basis as well.

\subsubsection{Sampling with monomers}

We will also be interested in the distribution of two monomers in the RVB states. In the case of the CDM, the distribution function of
the monomer separation can be measured just by keeping track of the two defects,\cite{Adams,Sandvik-Moessner} but in the RVB we have to explicitly
introduce two monomers by removing both spins on a randomly chosen valence bond which is common to both the ket and bra bond configurations. 
Note that valence bond states with monomers are orthogonal unless the monomers are at the same locations in both states. We use the loop 
algorithm to sample the bond configuration space, and periodically we also move the monomers. Such a move can be done in combination with 
the move of a valence bond that is common to the two states, as shown in Fig.~\ref{update}(c). This can always be accepted if there is no change in the bond 
length (one could also consider updates where a monomer moves and a bond length
changes, which we do not do here). We update the position of two monomers 
in turn after each sweep of 
bond updates, when possible, and measure the distribution probabilities $M({\bf r})$ as a function of distance ${\bf r}$ between the two monomers. 

Note that if we assign spins to the monomer the situation is different, due to the overcompleteness of the basis. In a system with, e.g., two
unpaired $\up$ spins, these two spins do not have to be located at the same sites in the ket and bra state---for a non-zero overlap it is only
required that they are pairwise connected by valence bonds in the transition graph (which now contains two broken loops with open ends terminated
by the unpaired spins). Such states with unpaired spins should be related to spinons,\cite{Read} but we will not pursue studies of them here. 
Valence bond states including unpaired spins have recently been studied in different systems.\cite{Wang10,Banerjee10}

\subsection{Winding numbers}
\label{windings}

A two-bond update cannot bring the system from one topological winding number sector to another, while the loop update can. In the case of the 
RVB, there are winding numbers both for the bra and the ket state, and because of the non-orthogonality of the basis these winding numbers can 
be different. We denote the full winding number of a configuration in this case as $W=(W_x^\alpha,W_y^\alpha;W_x^\beta,W_y^\beta)$.
In a grand canonical ensemble of all winding numbers, the sectors have different weight, which can be computed using Monte Carlo sampling
with the loop updates simply by keeping track of the number of configurations generated in each sector. Results for such weights are presented 
below in Sec.~\ref{sec:Wprob}.

The loop algorithm for the CDM remains ergodic in the grand-canonical winding-number space even for very large systems, i.e., the loops can 
easily become very long and span the system. These long loops are related to deconfined monomers.\cite{monomernote} The RVB simulations, in the 
case of short-bond states, in practice become stuck in some fixed winding-number sector for large $L$. However, the shortness of the RVB loops 
does not imply monomer confinement, as these loops are not directly related to states with monomers.\cite{monomernote} The loops for short-bond 
two-dimensional RVB states are typically very short (rarely exceeding $12$ bonds in the case of the length-$1$ bonds only). This results in 
rather large error bars for computed quantities for $L\agt 50$, seen in grand-canonical results to be discussed further below. In practice, 
for large systems we will therefore study canonical ensembles in different fixed winding number sectors. Starting with a configuration 
initially prepared with a desired winding number (such as those illustrated in Fig.~\ref{sectors}), two-bond updates explicitly conserve 
the winding number while loop updates in practice do as well, for large systems within reasonable simulation times.

\section{RESULTS}
\label{results}

The ground state of the QDM at the RK point is the equal amplitude superposition of classical dimer states. The CDM can therefore give some insights 
into properties of the RVB system as well, as long as the non-orthogonality of the valence-bond basis (i.e., the internal singlet structure of the
valence bonds of the RVB) does not play an important role.\cite{RK1} The quantitative validity of this approach is tested here by comparing the properties 
of the CDM and the short-bond RVB state. We present the winding number distributions of both models in Sec.~\ref{sec:Wprob}, then briefly discuss the 
standard spin correlation function of the RVB in Sec.~\ref{spincorrs}. In Sec.~\ref{dimerdimer} we study the four-spin VBS correlation function Eq.~(\ref{dimer}) 
of the RVB (which we also refer to as a dimer-dimer correlation function) and compare with analogous results for the well known dimer-dimer correlations of 
the CDM. In this section we consider the winding number sector $W=(0,0)$ and later, in Sec.~\ref{nonzerow}, discuss also correlations in systems with 
nonzero winding number. In Sec.~\ref{monomerdist} we study the monomer distribution functions and in Sec.~\ref{longbond} systems including the longer bonds.

\subsection{Sector probabilities}
\label{sec:Wprob}

\begin{figure}
\centerline{\includegraphics[angle=0,width=7.5cm, clip]{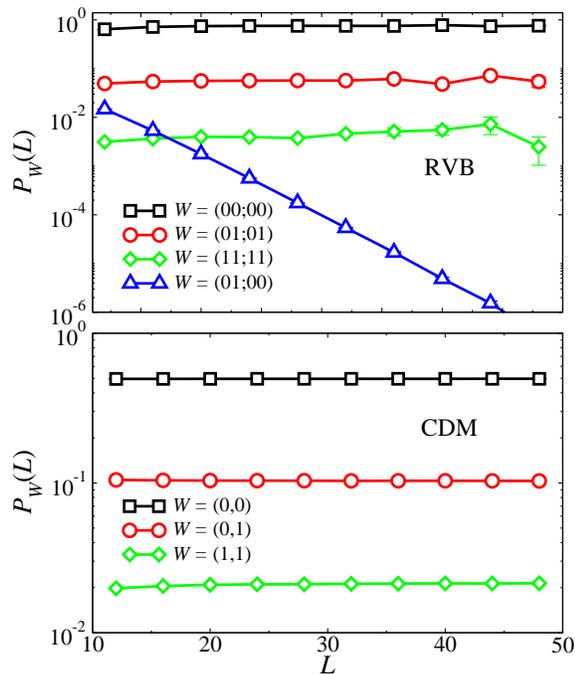}}
\caption{(Color online) Winding number probabilities obtained in simulations with the loop algorithms for the RVB and CDM (with only the 
shortest bonds, as in Fig.~\ref{update}). Results for several low-$W$ sectors of the CDM (lower panel) and RVB (upper panel) are shown versus 
the lattice size on a lin-log scale. In the RVB, the probability of the off-diagonal sector $W=(0,1;0,0)$ vanishes exponentially with $L$, 
reflecting the orthogonality (when $L\to \infty$) of states in different winding number sectors.}
\label{Wprob}
\end{figure}

We simulated the grand-canonical ensemble of winding numbers, as explained in Sec.~\ref{windings}, and accumulated the probabilities of several 
different sectors as shown in Fig.~\ref{Wprob}, for both the RVB and CDM, and for various system sizes $L$. The $W=0$ [$(0,0)$ for the CDM and
$(0,0;0,0)$ for the RVB) sector is dominant in both cases, with the probabilities in the higher-$W$ sectors decreasing rapidly. The probabilities of 
these low-$W$ sectors clearly converge to $L$-independent non-zero constants, rapidly with $L$ for the CDM, and also for the diagonal ($W^\alpha=W^\beta$) 
sectors of the RVB (although the RVB data are much noisier for the large systems). By contrast, the probabilities of the off-diagonal sectors of the RVB, 
here exemplified by $W=(0,1;0,0)$, decay exponentially to zero, which reflects the expectation that the states in different winding number sectors 
should become orthogonal in the thermodynamic limit.\cite{Bonesteel} In the following, when considering winding number sectors of the RVB we will
focus on the diagonal sectors and for simplicity denote the total winding number by $W=(W_x,W_y)$ in the same way as for the CDM.

\subsection{Spin correlations in the RVB state}
\label{spincorrs}

\begin{figure}
\centerline{\includegraphics[angle=0,width=7.5cm, clip]{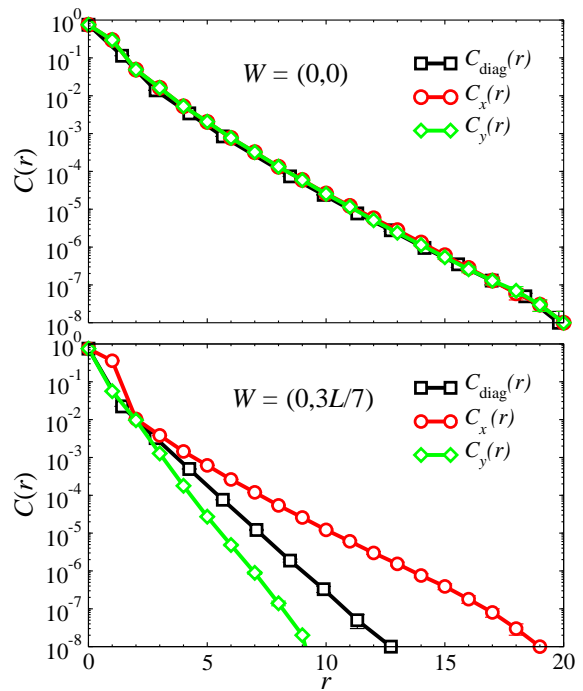}}
\caption{(Color online) Spin correlations versus lattice distance $r$ in the short-bond RVB in the sector of winding numbers $W=(0,0)$ 
(top panel) and $W=(0,3L/7)$ (bottom panel) computed using $L\times L$ lattices with $L=48$. Results are shown for the separation $(x,y)$ 
taken along the two axis, $(r,0)$, $(0,r)$, as well as on the diagonal, $(r/\sqrt{2},r/\sqrt{2})$.}
\label{spincorr}
\end{figure}

The spin-spin correlation function of the RVB has been studied before and is 
known to decay exponentially for a 2D system with short bonds (while
a system with sufficiently slow decay of the probability of long bonds has long-range antiferromagnetic order).\cite{Liang,Beach2} Here, we only comment
briefly on the role of the winding number. For unequal $x$ and $y$ winding numbers, $W_x\not =W_y$, the CDM and RVB systems clearly do not have the 
$90^\circ$ rotational symmetry of the square lattice. We will investigate the directional dependence of the four-spin dimer-dimer correlations below. 
Here, in Fig.~\ref{spincorr}, we show results for the spin-spin correlations in two different winding number sectors. The correlations are always 
exponentially decaying with distance, with a faster decay in the same direction as the one in which a non-zero winding number is imposed.

\subsection{Dimer Correlations}
\label{dimerdimer}

In the CDM, the dimer-dimer correlation function $D_{xx}({\bf r})$ is defined in the standard way using the bond occupation number $n_x(i)=0,1$ on the 
link of the lattice between site $i$ and its neighbor at distance $(1,0)$; $D_{xx}({\bf r}_{ij})=\langle n_in_j\rangle$. The four-spin correlation
function Eq.~(\ref{dimer}) of the RVB instead involves the loop estimator 
Eq.~(\ref{scorr4}). This reduces to the CDM form for SU($N$) spins when $N\to \infty$ 
and the basis becomes orthogonal [in the representation of SU($N$) in which the factor $1/2$ in the off-diagonal matrix element in Eqs.~(\ref{cijop1}) 
and (\ref{cijop2}) is replaced by $1/N$;\cite{Read2} see, Ref.~\onlinecite{Beach1} for computations with such basis states]. For $N=2$, considered here, 
significant differences between the RVB and CDM can be expected.

\begin{figure}
\centerline{\includegraphics[angle=0,width=7.5cm, clip]{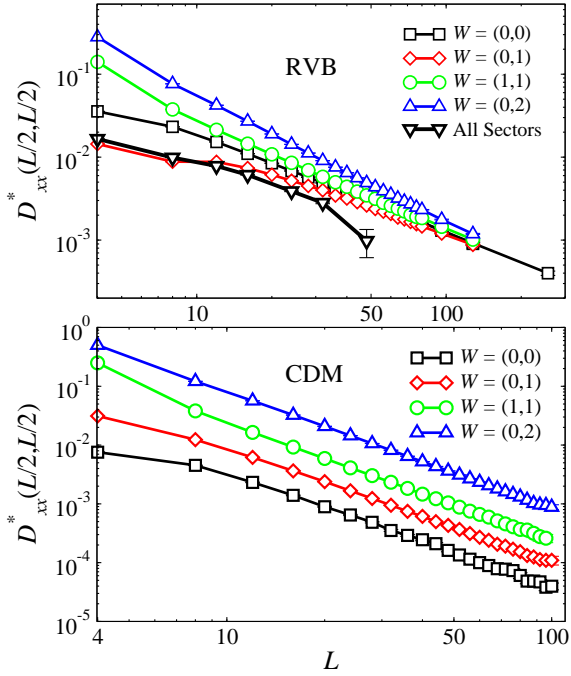}}
\caption{(Color online) Dimer-dimer correlation function difference Eq.~(\ref{dstar}) at the maximal distance versus the lattice size. The upper panel shows 
results for the quantum RVB in different topological sectors as well as in the grand canonical ensemble (including all winding number sectors, in which case 
the fluctuations between sectors becomes very slow for large systems, as reflected in the large error bar for $L=48$). All correlations converge to the same 
power-law decay as system size increases. The power, based on the $W=(0,0)$ data for large $L$, is $\alpha = 1.191(6)$. The lower panel shows results 
for the CDM, which are consistent with $\sim 1/r^2$ (shown with the solid line) for all winding number sectors.} 
\label{crr}
\end{figure}

Since we are using periodic boundary conditions, the maximal separation to be used in the correlation function is $(L/2,L/2)$ on a $L \times L$ lattice. 
We first investigate the dominant part of the correlation function, which in the CDM is a mixture of a staggered component, at $q=(\pi,\pi)$ in reciprocal 
space, and columnar correlations, at $q=(\pi,0)$ and at $(0,\pi)$.\cite{Stephenson} The asymptotic decay of these correlations can be accessed through the 
difference between the real-space correlations at two distances, e.g.,
\begin{equation}
D^*_{xx}(x,y)= D_{xx}(x,y)-D_{xx}(x-1,y).
\label{dstar}
\end{equation}
This quantity at the longest distance ${\bf r}=(L/2,L/2)$ is graphed versus $L$ in Fig.~\ref{crr} for both the RVB and the CDM in several fixed winding 
number sectors. 

For the CDM, the decay with $L$ is consistent with the known $\sim 1/r^2$ decay of the dominant correlations. Apart from an overall prefactor 
that depends on the winding number, there are only minor differences between the different winding sectors for small systems. 
The dependence of the results on the winding number is stronger for the RVB, but, as expected, also here the exponent $\alpha$ in the power-law form
$1/r^\alpha$ becomes independent of $W$ for large $L$ (as long as the relative winding number $W/L \to 0$ when $L \to \infty$).
Unlike the CDM, in this case the prefactor of the power-law form also converges as 
$L \to \infty$, i.e., the correction to the prefactor decays as some power higher than $\alpha$. 

In Fig.~\ref{crr}, we also show results in the grand-canonical winding number 
ensemble, which, as discussed in Sec.~\ref{windings}, suffers from problems with non-ergodic sampling for $L \agt 50$ (reflected in the large error bar for 
$L=48$). For extracting the asymptotic form of the correlations, the $W=(0,0)$ sector is the best choice and gives $D(r) \propto 1/r^{\alpha}$ with 
$\alpha = 1.191(6)$ for large systems. While the behavior is, thus, qualitatively similar to the CDM, the exponent differs considerably. The reduced value
of the exponent can be interpreted as the RVB state being closer to an ordered VBS than might have been anticipated based on the known CDM dimer correlations.

There are two sources of differences between the correlations in the CDM 
and the RVB: the form of the estimator Eq.~(\ref{scorr4}) as well as 
the weighting of the bra and ket valence bond states with the 
loop factor $2^{n_{\alpha\beta}}$ for the RVB instead of the equal 
superposition of the individual bond configurations in the CDM. 
We have also measured the dimer correlations of the RVB in the same 
way as in the CDM, by just using the bond occupation numbers in the bra and the 
ket states (but with the correctly weighted sampling of the RVB). We find the 
same exponent $\alpha \approx 1.20$ as above, 
which shows that the source of the 
different power-law is only the different weighting of the states. This could 
also have been anticipated based on the fact that the 
spin-spin correlation function of the RVB is exponentially decaying, 
which translates into short loops in the transition 
graph.\cite{Sutherland} The loop estimator Eq.~(\ref{scorr4}) of the 
four-spin dimer correlation function is therefore still local and 
cannot change a power law.

\begin{figure}
\centerline{\includegraphics[angle=0,width=8cm, clip]{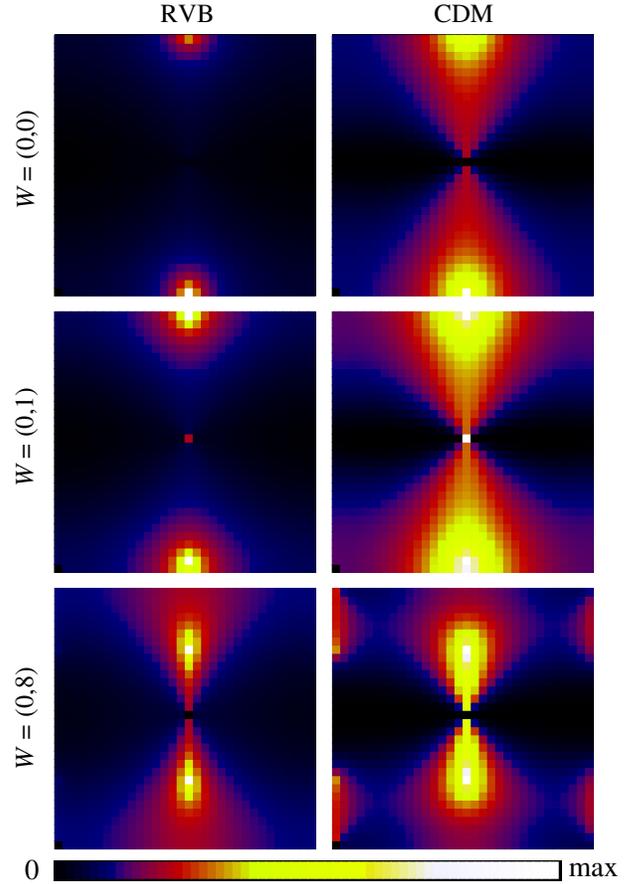}}
\caption{(Color online) Fourier transform ${\bf S}({\bf q})$ of the dimer-dimer correlation function $D_{xx}({\bf r})$ for systems of size $L=32$. The squares 
represent the full reciprocal space $q_x,q_y \in [0,2\pi]$. Results in winding number sectors $W=(0,0)$, $W=(0,1)$, and $W=(0,8)$ are shown for the RVB (left) 
and CDM (right). The location of the broad (``incommensurate'') peak in both cases is ${\bf Q}=(\pi,2\pi W_y/L)$. The sharp peak at $(\pi,\pi)$ is due to a 
nonzero average staggered dimer order induced by a nonzero winding number. This peak has been removed in the graphs $W=(0,8)$ in order to make the other 
features of the correlations better visible. 
The height of the peaks as a function of the system size is analyzed in Fig.~\ref{diffw}.}
\label{2dfft}
\end{figure}

\begin{figure}
\centerline{\includegraphics[angle=0,width=8cm,clip]{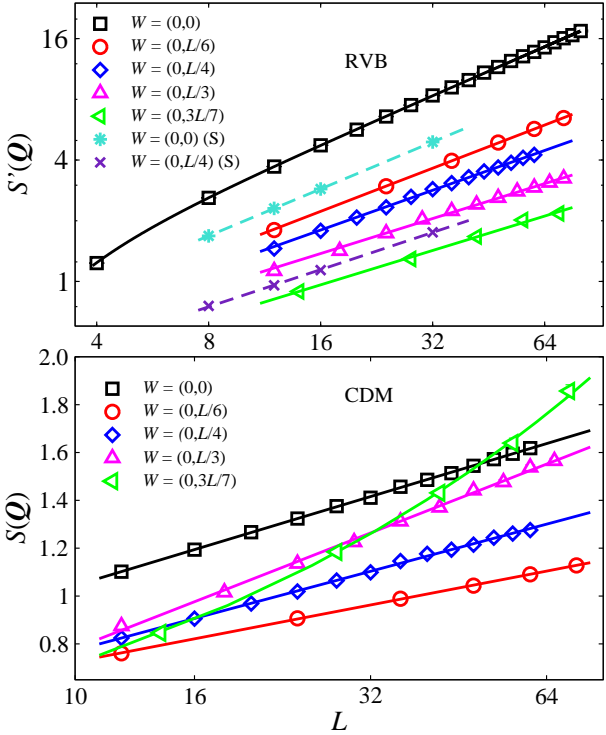}}
\caption{(Color online) Peak values of the dimer structure factor, where ${\bf Q}=(\pi,2\pi W_y/L)$, versus the system size in sectors with 
different winding $W_y$. The modified definition $S'({\bf Q})$ for the RVB is given in Eq.~(\ref{eq:SQ}). Note the different $y$-axis scales used for 
the two models (logarithmic for the RVB and linear for the CDM). In the CDM (lower panel) the behavior is consistent with a log divergence (as
shown with fitted lines) for small winding numbers, but for larger $W$ it appears that the behavior is instead governed by a power law (which then may 
be the case for all $W_x/L>0$ for sufficiently large systems). The curve through the $W=(0,3L/7)$ data shows $S({\bf Q}) \propto L^{0.48}$. In the RVB 
(upper panel) the exponent of the power-law divergence decreases slightly with increasing winding number. The legends with $(S)$ correspond to
the peak values of the full structure factor $S({\bf Q})$.} 
\label{diffw}
\end{figure}

The Fourier transform of the full dimer-dimer correlation function $D_{xx}({\bf r})$ is the structure factor $S({\bf q})$. This quantity gives 
a more detailed picture of the long-distance behavior of the dominant correlations. Representative results for the $S({\bf q})$ for $L=32$ systems in three different 
winding number sectors $(0,W_y)$ are shown in Fig.~\ref{2dfft}. In this section we focus on the $W=(0,0)$ sector and leave discussions of nonzero 
winding numbers to Sec.~\ref{nonzerow}. The ``bow-tie'' feature seen for $W=(0,0)$ in the CDM is well known and understood based on the mapping of the 
system to a height model (see Appendix~\ref{app:height}). The system has two kinds of power-law correlations: an effectively dipolar kind, which is 
responsible for the ``pinch-point'' singularity at $\qq=(\pi,\pi)$ (see Sec. \ref{sec:app:dimer-dipolar}), and a ``critical'' kind with variable 
exponents, which leads to a broad peak at ${\bf Q}=(\pi,0)$ diverging logarithmically with the system size, as shown in the lower panel of Fig.~\ref{diffw}. 
In the RVB the peak is much sharper and diverges faster, as a power law (as shown in the upper panel of Fig.~\ref{diffw}) on account of the 
real-space form $1/r^\alpha$ with $\alpha \approx 1.2<2$ of the dimer correlation function.

When the Fourier transform $S({\bf q})$ is computed post-simulation based on all computed real-space correlations, the measurements in the simulations 
are expensive, requiring $\mathcal{O}(N^2)$ operations to take full advantage of spatial averaging. In the CDM, we can instead easily just compute $S({\bf Q})$ 
at the single wave-vector ${\bf Q}$ directly in the simulations at a much lower cost of $\mathcal{O}(N)$ to access larger system sizes. In the RVB, this 
speed-up is not possible, however, because we are there really measuring a four-spin correlation function that cannot be simply expressed as a product 
of two-spin correlators, as discussed in Appendix A, and there is no obvious way of avoiding the $\mathcal{O}(N^2)$ scaling of this measurement.

In order to have a similar quantity, which scales with the system size in the same way as $S({\bf Q})$ but for which the measurements require only 
$\mathcal{O}(N)$ operations, we define a modified structure factor $S'({\bf Q})$ for the RVB as 
\begin{equation}
S'({\bf Q})= \langle \widetilde{B}_{x}^*({\bf Q})\widetilde{B}_{x}({\bf Q}) \rangle
\label{eq:SQ}
\end{equation}
where $\widetilde{B}_{x}({\bf Q})$ is the Fourier transform of the spin-spin correlator matrix element $\langle V_\beta |({\bf S}_i \cdot {\bf S}_j)
|V_\alpha\rangle$ for an individual configuration in the RVB simulation (i.e., obtained from a transition graph, which gives values $\in \{-3/4,0\}$ for 
each nearest-neighbor bond on the lattice). This definition of the peak value differs from the full Fourier transform $S({\bf Q})$ of the four-spin 
dimer correlator $D(r)$, essentially because it does not contain any information on the order of the site indices in the matrix element 
$\langle V_\beta |({\bf S}_k \cdot{\bf S}_l)({\bf S}_i \cdot {\bf S}_j)|V_\alpha\rangle$, which plays a role in the transition-graph two-loop 
estimator of the dimer correlation function (as discussed in Appendix A). In particular, the modified quantity misses certain negative contributions 
arising in some cases where all four indices belong to the same loop 
[see Eq. (\ref{scorr4})]. Therefore, we expect $S'({\bf Q}) > S({\bf Q})$, which is also confirmed by 
results for both quantities in small systems, as shown in the upper panel of Fig.~\ref{diffw}. The form of the power-law divergence is the same, however.

Overall, there is significant directional dependence in the dimer correlations, but for $W=(0,0)$ the RVB results in Fig.~\ref{diffw} confirm that the peak at 
$(\pi,0)$ (corresponding to columnar-modulated correlations) is sufficiently isotropic for the size dependence of the Fourier peak to be directly related 
to the exponent of the power-law decay $1/r^\alpha$ found above for the real space correlation (and, it should be pointed out, the exponent $\alpha$ also 
comes out consistently to the same value when extracted in different directions in real space). 

With $S'({\bf Q})$ diverging with the system size $L$ 
as $L^{\alpha_Q}$, we expect $\alpha_Q \approx 2-\alpha$ and the data confirm this. For instance, the $W=(0,0)$ data in the upper panel 
of Fig.~\ref{diffw} was fitted to a function $f(L)=b_QL^{\alpha_Q}+b_2L^{\alpha_2}$,
where $\alpha_2 < \alpha_Q$ (and typically also $\alpha_2 < 0$) and this 
correction term is added in order to include data for the full range of 
systsem sizes. By using this form we obtained $\alpha_Q=0.800(2)$, which is 
in good agreement with $\alpha=1.191(6)$ but with a smaller error bar. 
Our best estimate for the exponent is, thus, $\alpha=1.200(2)$. 
Here the error bar is purely statistical and there may still be some 
systematical errors present as well (likely of the same order), 
arising from neglected higher-order corrections. 

\subsection{Correlations with nonzero winding number}
\label{nonzerow}

\begin{figure}
\centerline{\includegraphics[angle=0,width=7.5cm, clip]{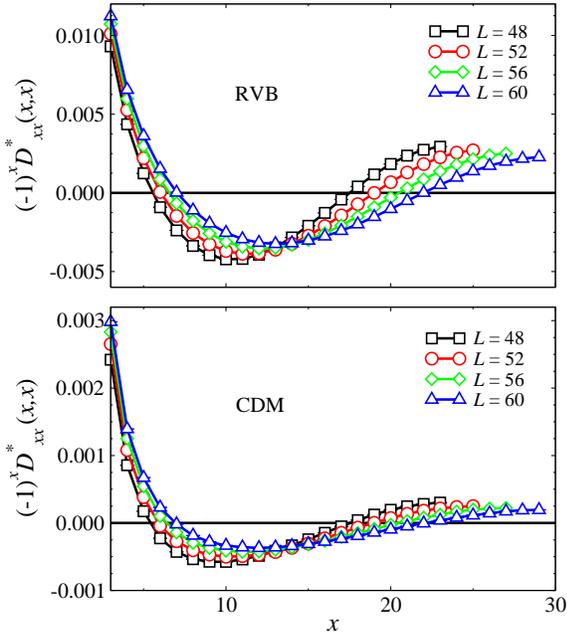}}
\caption{(Color online) Dimer correlation differences versus $x$ [where the separation ${\bf r}=(x,x)$] along the diagonal lattice direction for systems of 
different size. The winding number is $W=(0,2)$, and therefore two phase shifts are seen (corresponding to a total of four domains). Note that the overall 
magnitude of these correlations is much larger in the RVB (upper panel) than in the CDM (lower panel).} 
\label{diag}
\end{figure}

In a background of nonzero winding number, all dimer-dimer correlations
should become modulated by the factor $\cos(\delta \QQ\cdot \rr)$, as derived 
using the height-model formalism in Appendix~\ref{app:height} and
shown explicitly as Eq.~(\ref{eq:Dxxcrit-modulated}), where $\delta\QQ= {2\pi}(W_x,W_y)/L$. 
Such a modulation is visible in the real-space dimer 
correlation function, as shown in Fig. \ref{diag} for $D^*_{xx}({\bf r})$ along the diagonal lattice direction, ${\bf r}=(x,x)$, for 
systems of different size with winding number $W=(0,2)$.
This implies that when $\rr$ is followed along the $[1,\pm 1]$ direction
through an entire period,
$2 (W_x \pm W_y) $ nodes of $D_{xx}(\rr)$ are crossed; indeed, 
Fig.~\ref{diag} for $W=(0,2)$ shows two changes of sign between $x=0$ and $L/2$, 
in both the CDM and the RVB cases.

\begin{figure}
\centerline{\includegraphics[angle=0,width=6.5cm, clip]{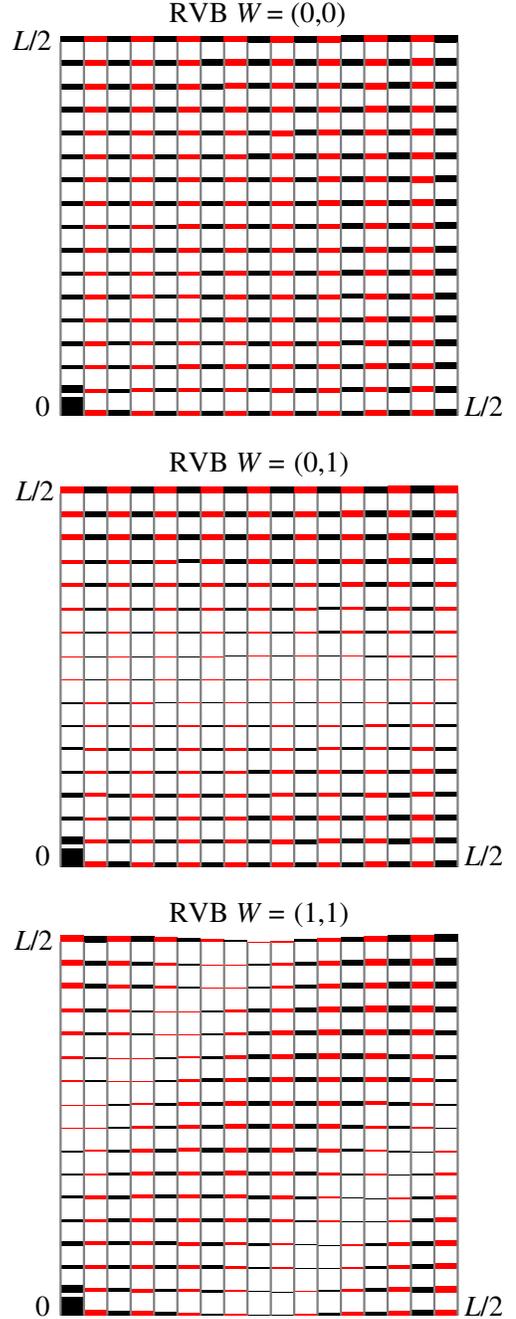}}
\caption{(Color online) Correlation patterns obtained from the dimer correlator $D_{xx}(x,y)$ by subtracting a constant and dividing the result by 
the leading power-law form $r^{-\alpha}$ (with $\alpha =1.2$ for the RVB and $\alpha=2$ for the CDM). The $(\pi,\pi)$ contribution was also removed
for the $W\not=0$ sectors (by going to Fourier space as in Fig.~\ref{2dfft}). Black and red (gray) bars represent positive and 
negative values (i.e., stronger and weaker dimer correlations), respectively. In the $W=(0,0)$ sector, 
a dominant columnar pattern is visible, while in the $W=(0,1)$ sector the correlations shift from weak-strong weak-strong to strong-weak strong-weak over a window of 
distances $\propto L$, corresponding to two nodal lines as stated in text. The origin is at lower left corner, and one quadrant ($L/2 \times L/2$) 
is shown of the possible separations. In the $W=(1,1)$ sector, correlations shift twice in a row, corresponding to the presence of two pairs of 
nodal lines.}
\label{2dreal}
\end{figure}

The correlation function $D_{xx}(x,y)$ in the full 2D space is shown for the RVB in Fig. \ref{2dreal}, where an overall background constant representing 
$D(r \to \infty)$ has been subtracted from $D({\bf r})$ and the remainder has been multiplied by $r^{\alpha}$ to make the modulations visible. An over-all
non-decaying staggered contribution present when $W\not=0$ has also been subtracted (see further discussion of this below and in Fig.~\ref{2dfft}).
The color coding shows positive and negative correlations, and the width of bars represent the magnitude of the correlations. In the winding number 
$W=(0,0)$ sector, the positive and negative values alternate in rows, showing that the overall dominant 
correlations are of columnar type. In the $W=(0,1)$ sector, a phase shift occurring around at $y=L/2$ is clear. The region over which the shift takes place 
is itself of size $\mathcal{O}(L)$, as expected since the amplitude is modulated proportional to a sine wave (which can be considered as a highly fluctuating 
critical delocalized domain wall). The results for the $W=(1,1)$ sector confirm the existence of two such delocalized nodes along the diagonal direction.
A similar pattern of phase shifts in the correlation function is seen in the CDM case as well, but is much weaker because of the significantly faster
decaying correlations (as is also clear in Fig.~\ref{diag}). 

To our knowledge, these correlations in sectors of fixed non-zero winding number have not 
been studied in detail previously (but were pointed out also in the parallel work by Albuquerque and Alet \cite{albu10}). In Appendix \ref{app:height}, we 
extend the height-model approach to this case as well (in  Sec.~\ref{app:winding-corr}). Here we only briefly discuss some of the main features, with 
the aim of comparing the RVB and CDM systems. 

Turning back to the Fourier space plot, Fig.~\ref{2dfft}, it includes representative results for the structure factor in three different 
winding number sectors $(0,W_y)$. Once the winding number is non-zero, it is clear that there is, for both models, a $\delta$-function peak 
in $S({\bf q})$ at $(\pi,\pi)$, reflecting a non-zero static staggered order parameter. Since this peak grows in proportion to the 
winding number, we have subtracted it off in some cases in Fig.~\ref{2dfft} to make the other features better visible. 

There are two notable features of these results, for both the RVB and CDM: (i) the pinch-point remains at $(\pi,\pi)$ and (ii) the singularity at 
$(\pi,0)$ present for $W_y=0$ is offset to $\QQ=(\pi,2\pi W_y/L)$,
which when $L\to \infty$ can be considered as an incommensurate peak at $\QQ\equiv(\pi,w)$, $w\in [0,\pi]$. This is exactly as expected 
from Eq.~(\ref{eq:delta-Q}) obtained within the height-model representation in Appendix~\ref{app:height}. Figure \ref{diffw} shows the system 
size dependence of the singular peak for different {\it large} winding numbers $W_y \propto L$. These features have been qualitatively expected in 
the case of the CDM based on several previous works~\cite{Zeng,Jacobsen,Fradkin1} (as outlined in Appendix~\ref{app:height}), but they are still
interesting to study quantitatively and to elucidate the similarities and differences between the CDM and RVB. It is already clear from Fig.~\ref{2dfft} 
that the divergence of the incommensurate peaks is much stronger for the RVB than the CDM, which is anticipated based on our result for the slow 
real-space decay of the dimer-dimer correlations in the RVB. 

For non-zero winding number, the correlations become significantly anisotropic, but we have not attempted to study their full functional form in real 
space or Fourier space. The exponent governing the asymptotic power-law decay is, however, expected to be direction independent, as discussed
in Appendix \ref{app:height}. The results in Fig.~\ref{diffw} indicate that $S({\bf Q})$ has the form $L^{\alpha_Q}$, with a weak dependence of the 
exponent $\alpha_Q$ on the location of the peak (i.e., the winding number), also as expected based on the height-model results in 
Appendix \ref{app:aniso}.

The incommensurate peak of the CDM was discussed by 
Fradkin et al.,\cite{Fradkin1} who pointed out a set of 
critical points in extended QDMs with more complicated diagonal and 
off-diagonal terms than the standard RK nearest-neighbor bond-pair 
interactions. The critical points extend from the conventional RK point 
at zero winding number, forming a complex fractal curve 
with devil's staircase features (forming a Cantor set). 
This critical curve separates a staggered dimer phase from one with a 
complex bond pattern with a large unit cell, which depends on the 
winding number. Similar transitions with a series of different VBS 
phases were studied in Ref.~\onlinecite{Fradkin2}. 
Our CDM results in Fig.~\ref{diffw} for large winding 
numbers suggest that the incommensurate peak may become power-law divergent (i.e., stronger than the logarithmic divergence obtaining 
at zero winding number). This is seen most clearly in the $W=(0,3L/7)$ graph, where it is clear that the divergence with $L$ is 
faster than logarithmic. A power-law fit, $L^{\alpha_Q}$ with $\alpha_Q=0.48(3)$ describes the data well. This is expected in the height 
scenario, since a nonzero background $W/L$ changes the effective stiffness to $K'$ as given by Eq.~(\ref{eq:Keff}). The exponent $\alpha$ 
of real-space correlations accordingly changes from $2$ and consequently the integral of $1/r^\alpha$ (the structure factor) should
diverge faster than logarithmically.

\subsection{Monomer distribution}
\label{monomerdist}

\begin{figure}
\centerline{\includegraphics[angle=0,width=7.5cm, clip]{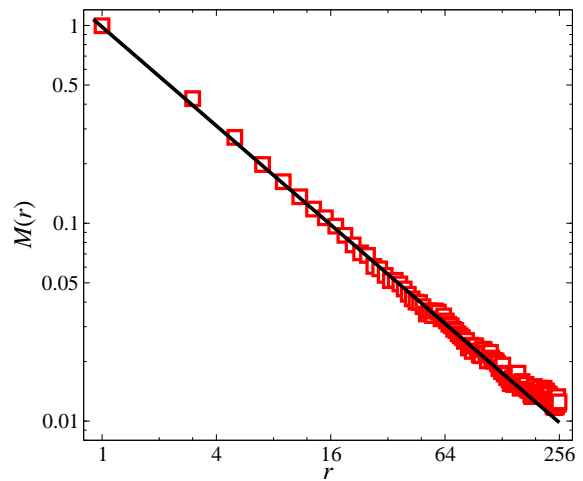}}
\caption{(Color online) Monomer distribution function in the RVB state on an $L=512$ lattice. The straight line is a fit to
the power-law form $1/r^{\beta}$ with ${\beta} = 0.830(9)$.}
\label{monomer}
\end{figure}

Monomers are expected to be deconfined in RVB states,\cite{Anderson} which provides an intuitive picture of spin-charge separation. Here we will
study two monomers in the RVB. It should be noted, however, that these monomers are bosonic, and hence the results cannot be directly related to a hole-doped 
RVB spin liquid. In that case the monomers should be fermions and, as discussed, e.g., in Ref.~\onlinecite{Read}, the sign rule we use here for the valence
bonds would have to be replaced by more complex signs. It is nevertheless interesting to compare the monomer-doped RVB and CDM systems considered as different 
statistical mechanical systems.

\begin{figure}
\centerline{\includegraphics[angle=0,width=7.5cm, clip]{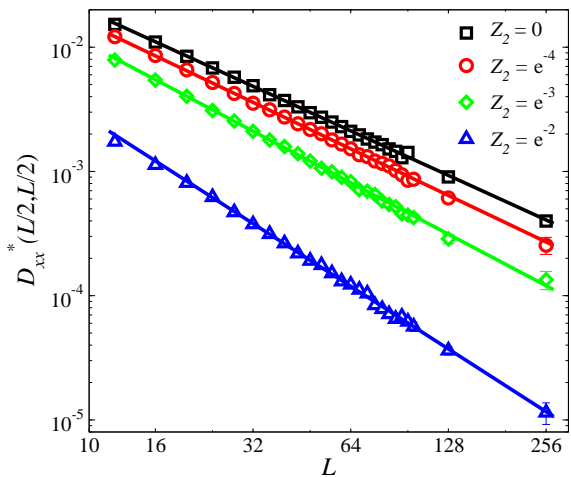}}
\caption{(Color online) Dimer-dimer correlation function difference $D_{xx}(r)$ for RVB systems in the $W=(0,0)$ sector with different fugacities $Z_2$ of long 
(fourth-neighbor) bonds (with the short-bond fugacity $Z_1=1$). The decay exponents grows with the long-bond fugacity. The values are 
given in Table~\ref{tab:exponents}.} 
\label{lbcrr}
\end{figure}

The monomer-monomer distribution function of the CDM is defined using the monomer density $m({\bf r}_i) =0,1$;
\begin{equation}
M({\bf r}_{ij})=\frac{\langle m({\bf r}_i)m({\bf r}_j)\rangle}{\langle m({\bf r}_i)m({\bf r}_i+\hat {\bf x})\rangle},~~~~
{\bf r}_{ij}={\bf r}_i-{\bf r}_j,
\label{monodef}
\end{equation}
where the normalization with the correlation at distance $r=1$ is a convention 
which makes it easy to compare results for different system sizes 
(i.e., results for fixed $r$ converge to a non-zero number with increasing 
size, even if the monomers are deconfined). It is known \cite{Stephenson} that 
this function for the short-bond CDM decays as $M(r) \propto 1/r^{\beta}$ with 
$\beta = 1/2$. This slow decay reflects monomer deconfinement, 
i.e., the function $\langle m({\bf r}_i)m({\bf r}_j)\rangle$ without 
the normalization in Eq.~(\ref{monodef}) decays to zero for 
fixed ${\bf r}_{ij}$ when $K \to \infty$. We use exactly the same 
definition of $M({\bf r})$ for the RVB, applying the procedures 
discussed in Sec.~\ref{algorithm} to sample 
monomer configurations (while in the CDM the loop algorithm for the 
bond sampling without monomers gives the monomer distribution function as 
a by-product \cite{Sandvik-Moessner,Adams}). 
Note that the winding number is not well defined in the presence of monomers, 
since they are associated with ``broken loops'' in the transition graph in Fig.~\ref{winding}.

The exponent $\beta=1/2$ for the CDM has been confirmed previously in Monte Carlo simulations on large lattices.\cite{Sandvik-Moessner} 
Figure \ref{monomer} shows our results for the RVB, 
using a system of size $L=512$ (for which the results for moderate 
separation of the monomer are sufficiently converged to 
extract the decay exponent). We find that the exponent $\beta \approx  0.83$ is significantly larger than in the CDM. The monomers are, thus, more 
strongly correlated to each other than in the CDM, but still deconfined. Note that in a long-range ordered VBS one would expect the monomers to be
confined.

\subsection{Including longer bonds}
\label{longbond}

As the next step after investigating the extreme short-bond RVB, it is natural to think about the role of the longer bond in spin liquids and the 
classical dimer model. In the case of the CDM, introducing bonds between next-nearest neighbors on the square lattice leads to exponentially decaying
dimer correlations and monomer confinement,\cite{Sandvik-Moessner} as on a triangular lattice with only nearest-neighbor bonds.\cite{Moessner1} 
However, with only bipartite bonds, the behavior is qualitatively similar 
to the short-bond model (as long as the fugacity for longer bonds decays 
sufficiently rapidly with the length of the bonds).\cite{Sandvik-Moessner} The dimer correlations decay as $1/r^\alpha$ with $\alpha=2$ not changing as longer bonds are introduced, but the monomer 
exponent $\alpha$ decreases from $1/2$. 

\begin{figure}
\centerline{\includegraphics[angle=0,width=7.5cm, clip]{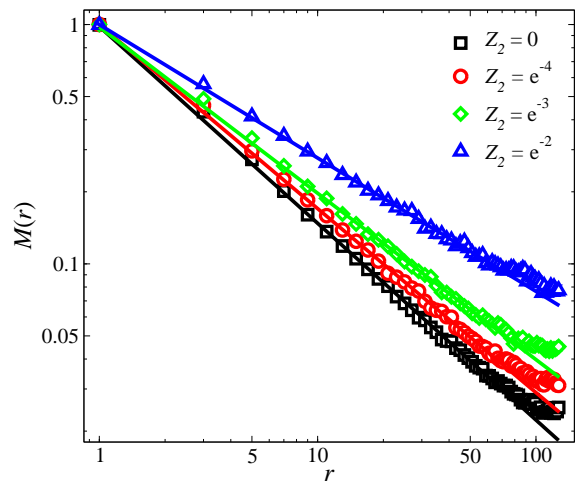}}
\caption{(Color online) Monomer distribution function $M(r)$ for RVB states with a small fraction of fourth-neighbor bonds on a lattice of size $L=256$. 
The straight lines are fits giving deconfinement exponents which decrease with increasing long-bond fugacity. The exponents are listed 
in Table~\ref{tab:exponents}.}
\label{lbmonomer}
\end{figure}

In the RVB, Marshall's sign rule cannot be applied if non-bipartite (frustrated) bonds are introduced. Due to the non-orthogonality of the basis,
there is, regardless of how signs beyond some simple Marshall rule are introduced, a sign problem in the Monte Carlo bonds sampling (due to non-positive 
definiteness of the state overlaps). We here study the effects of bipartite valence bonds connecting fourth-nearest neighbors, i.e., of ``shape'' $(x,y)=(2,1)$ 
and all symmetry-related shapes, as was done previously for the CDM.\cite{Sandvik-Moessner} We use small fugacities $Z_2=e^{-2}$, $Z_2=e^{-3}$ and $Z_2=e^{-4}$ 
for the longer dimers and $Z_1=1$ for the short bond connecting nearest neighbors. In the RVB, since we work with the amplitude product states 
[Eq. (\ref{hprod})], we just use the ``fugacities'' as another notation for the 
RVB amplitudes; $h(r=1)=Z_1=1$, $h(r=\sqrt{5})=Z_2$.

Spin correlations have been previously studied in the presence of long bonds, 
including exponential and power-law decays of the length-dependent fugacities.\cite{Liang,Beach2} Here we again focus on the dimer-dimer correlations and 
monomer distribution function.

The exponent of the dimer-dimer correlations changes with the fugacity of long bonds, as shown in Fig.~\ref{lbcrr} and Table~\ref{tab:exponents}. The change 
can be seen even more obviously in higher winding number sectors (not shown in the figure). Note also that the spin correlations increase when longer bond are 
introduced.\cite{Liang, Beach2} Fig.~\ref{lbmonomer} shows the monomer distribution $M(r)$ as defined in Eq.~(\ref{monodef}). Similar to the CDM,\cite{Sandvik-Moessner} 
the confinement exponent changes with fugacity of long bonds. The higher the fugacity of long bonds, the lower is the monomer deconfinement exponent.

\begin{table}[t]
\caption{Dimer-dimer and monomer exponents obtained for the CDM and RVB systems at different fugacities $Z_2$ for the next-shortest 
bonds (of length $\sqrt{5}$).}
\centering
\begin{tabular}{llll}
\hline\hline
~~Model~~~~~~~~~~~~ &$Z_2~~~~~~~~~~~~$ & $\alpha$~~~~~~~~~~~~~~~~~~~~ & $\beta$  \\ 
\hline
~~CDM &  0      & $1.98(1)$ & $0.4996(5)$~~\\
~~CDM & $e^{-4}$ & $2.17(2)$ &  $0.447(2)$~~\\
~~CDM & $e^{-3}$ & $2.44(8)$ &  $0.392(1)$~~\\
~~CDM & $e^{-2}$ & $2.7(2)$ &  $0.302(1)$~~ \\
~~RVB &  0      & $1.191(6)$  & $0.830(9)$~~\\
~~RVB & $e^{-4}$ & $1.255(5)$  & $0.775(5)$~~\\
~~RVB & $e^{-3}$ & $1.377(10)$  & $0.707(5)$~~\\
~~RVB & $e^{-2}$ & $1.676(12)$ & $0.563(6)$~~ \\
\hline
\end{tabular}
\label{tab:exponents}
\end{table}

\section{Height model interpretation}
\label{sec:height-anal}

All of the numerical results found in these simulations can be compared with results obtained in 
the framework of the ``height model'' introduced in Sec.~\ref{sec:height-intro} and elaborated
in appendix~\ref{app:height}.  According to that description, each of the following can be written as a 
function of a single parameter, the height stiffness $K$:
\begin{itemize}
\item[(1)] 
The sector probabilities $P(W_x,W_y)$ presented in Fig.~\ref{Wprob}.
\item[(2)] 
The exponent $\alpha$ of critical dimer
correlations, inferred from the $L$-dependence of
the structure factor at  $\QQ = (\pi,0)$ [the peak-value at winding number $W=(0,0)$ as shown in  Fig.~\ref{diffw}], and also from the $L$
dependence of these same correlations at $\rr=(L/2,L/2)$ in real space, as plotted in Fig.~\ref{crr}.
\item[(3)] 
The decay exponent $\beta$ of the monomer distribution function
$M(r)$ as presented in Fig.~\ref{monomer}.
\item[(4)] 
The coefficient of the ``pinch-point'' singularity in the structure factor $S(\qq)$ as shown in Fig.~\ref{2dfft}.
\end{itemize}

\noindent
We can use these relations to reduce the different results to independent estimates of the stiffness, which we call $K_P$, $K_\alpha$, $K_\beta$, and $K_S$,
from these respective measurements.  The agreement (to be demonstrated below) of these is powerful evidence that a height-like field 
theory underlies the RVB state.  That is well-known to be true for the CDM state, but the extension to the RVB is non-trivial, due to
the configuration space here consisting of two bond configurations weighted by their transition-graph loops, as discussed in Sec.~\ref{basis}. 
Indeed, we have not derived the height-model representation explicitly for the RVB. We will make some comments on the feasibility of actually
deriving the effective model below.

\subsection{Four ways to extract stiffness}

We now run through the ways in which we get four independent measurements of the height stiffness $K$.
CDM results are presented in parallel to the RVB  results, firstly to check the systematic errors in our fitting 
procedures against exactly known results, and secondly to emphasize the similar behaviors.

\subsubsection{Sector probabilities}

\begin{table}
\caption{Stiffness parameter $K_P$ in the infinite CDM and RVB systems inferred 
from the winding-number sector probabilities 
(from data in Fig.~\ref{Wprob}) according to Eq.~(\ref{kpequation}).}
\centering
\begin{tabular}{lllll}
\hline\hline
     & CDM & & RVB  &  \\
\hline
~$(W_x,W_y)$~~ & $P(W_x,W_y)$~  & $K_P$   &  $P(W_x,W_y)$ & $K_P$ \\ 
\hline 
~(0,0) &    0.49625(4)~   &    ---      & 0.764(5)  &  ~---       \\
~(1,0) &    0.10321(3)~  &  0.19628(3)~  & 0.057(2)  & ~0.325(5)  \\
~(1,1) &    0.02146(1)~  &  0.19629(3)~  & 0.0043(5)  & ~0.324(7)   \\
~(2,0) &    0.000925(2)~ &  0.19642(8)    & ---       &  ~---       \\ 
\hline
\end{tabular}
\label{tab:KP}
\end{table}

Table~\ref{tab:KP} gathers together the numerical sector probabilities from the data sets in Fig.~\ref{Wprob}. As seen in the figure, the smaller 
sizes show noticeable finite-$L$ corrections, which are expected to be $O(1/L^2)$ due to the quartic correction Eq.~(\ref{eq:f-quartic}) to the free energy 
density. The larger sizes show larger statistical errors particularly for the RVB case, as explained in Sec.~\ref{windings}. In order to partially 
account for finite-$L$ corrections of leading order and higher, which we need to extract the probabilities at $L\rightarrow \infty$ with relatively 
smaller statistical fluctuations by using a large set of lattice sizes, we use suitable polynomial fitting functions (some times without linear term) 
to extrapolate values in the thermodynamic limit.

\begin{figure}
\centerline{\includegraphics[angle=0,width=8.4cm, clip]{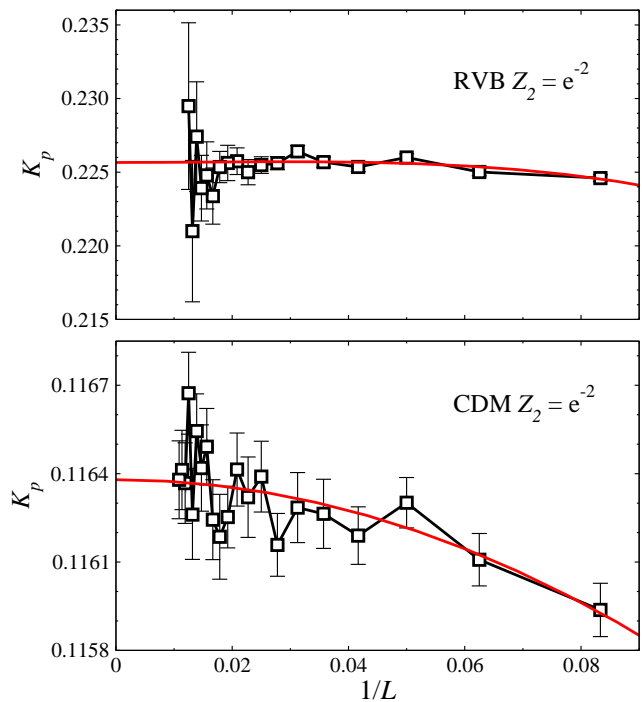}}
\caption{(Color online) $K_P$ value calculated in the $W=(0,1)$ sector according to Eq.~(\ref{kpequation}) for systems with fugacity $Z_2=e^{-2}$ for long bonds
and different lattice sizes. RVB and CDM results are shown in the upper and lower panel, respectively, as a function of the inverse system size 
$1/L$. The curves are second-order polynomial fits, not including the linear term.}
\label{stiffness}
\end{figure}

According to Eq.~(\ref{eq:P-W}), we expect $P(W_x,W_y) \propto \exp[-8 K(W_x^2+W_y^2)]$, and thus, we define
   \beq
       K_P\equiv -\frac{\ln[P(W_x,W_y)/P(0,0)]}{8 (W_x^2+W_y^2)}.
       \label{kpequation}
   \eeq
This expression clearly gives consistent results for every pair $(W_x,W_y)$, for either model as shown in Table~\ref{tab:KP}. The $K_P$ values in this table 
are calculated directly from the corresponding sector probabilities presented next to them. The $K_P$ values included in Table~\ref{tab:grand} are taken from 
the $W=(0,1)$ sector, as that has the smallest error bars (and also should be the best in terms of originating from a weak ``tilt'' field). As indicated 
by Fig.~\ref{stiffness}, the $K_P$ value does not depend much on system size $L$ for $L$ larger than $\approx 50$. Therefore, in order to obtain smaller 
statistical errors, we presented $K_P$ in Table~\ref{tab:grand} with the same method described above for extrapolating winding sector probabilities in 
the thermodynamic limit. As an example, polynomial fitting functions are shown in Fig.~\ref{stiffness}.

\subsubsection{Critical dimer correlations}

We have [Eq.~(\ref{eq:alpha}) in Appendix~\ref{sec:app:dimer-critical}] that
$\alpha = \pi/8K$; hence we define
   \beq
           K_\alpha \equiv \frac{\pi}{8 \alpha}.
   \eeq
The values of $\alpha$ summarized in Table~\ref{tab:exponents} could in principle all be obtained by fitting the size dependence of the 
peak-value $S({\bf Q})$ of the dimer structure factor, i.e., according to the peak-height analysis illustrated in Fig.~\ref{diffw} in the case of 
the RVBs. However, this approach requires a very significant computational effort for large lattices. We therefore use an easier but still 
reasonably accurate way to extract $\alpha$, by fitting the real-space long-distance dimer correlator $D^*_{xx}(L/2,L/2)$ as in Fig.~\ref{lbcrr} 
by a power-law [as expected according to Eq.~(\ref{eq:C-crit-decay})]. For non-zero $Z_2$ cases in the CDM, this approach does not work 
well, however, because $\alpha$ increases with the fugacity, becoming larger than $2$, and therefore the critical term is overshadowed by the 
stronger dipolar term (which always decays as $1/r^2$; see Sec.~\ref{sec:app:dimer-critical}) and is hard to detect. In contrast, in the RVB $\alpha < 2$ 
always and the critical term is dominant. A better way to find $\alpha$ in the 
CDM is to extract values by a fit of $|D^*_{xx}(x,x)|$ (along the diagonal
axis) for a range of distances $x$ on a large lattice, since the dipolar term vanishes on this axis. The corresponding $K_{\alpha}$ values are 
listed in Table~\ref{tab:grand}.

\begin{table}
\caption{Stiffness estimates obtained from the four kinds of measurements discussed in the text; $Z_2$ is the fugacity for dimers of length $\sqrt{5}$.}
\centering
\begin{tabular}{llllll}
\hline\hline
~Model &~~$Z_2$~~~& $K_P$~~~~~~~~~~~ & $K_\alpha$~~~~~~~~ & $K_\beta$~ &  $K_S$ \\ 
\hline
~CDM & ~ 0      &  0.19628(4)  & 0.198(1) & 0.1962(2)~ & 0.1959(7) \\
~CDM & ~$e^{-4}$ &  0.17547(4)  & 0.182(2) & 0.1755(8)~      & 0.1794(3) \\
~CDM & ~$e^{-3}$ &  0.15065(6)  & 0.161(5) & 0.1539(4)~      & 0.1582(4) \\
~CDM & ~$e^{-2}$ &  0.11638(3)  & 0.14(1)  & 0.1186(4)~      & 0.1234(1) \\ 
~RVB & ~ 0      &  0.323(5)    & 0.330(2) & 0.326(4)~  & 0.3242(4)\\ 
~RVB & ~$e^{-4}$ &  0.3067(8)   & 0.313(1) & 0.304(2)~ & 0.3081(2) \\
~RVB & ~$e^{-3}$ &  0.2774(5)   & 0.285(2) & 0.278(2)~ & 0.277(1)\\
~RVB & ~$e^{-2}$ &  0.2258(1)   & 0.234(2) & 0.221(2)~ & 0.22619(2) \\
\hline
\end{tabular}
\label{tab:grand}
\end{table}

\subsubsection{Monomer pair distribution correlations}

We have [Eq.~(\ref{eq:beta}) in Appendix~\ref{sec:app-monomers}]
that $\beta= 8K/\pi$; hence we define
   \beq
           K_\beta\equiv \frac {\pi \beta}{8}.
   \eeq
This quantity extracted from the exponents listed in Table ~\ref{tab:exponents},
where the values originate from fits to the $r$-dependence of the monomer 
distribution function (Fig.~\ref{lbmonomer} in the case of the RVBs), is listed in Table ~\ref{tab:grand}.

\subsubsection{Coefficient of the pinch-point in $S(\qq)$}

At $\QQ=(\pi,\pi)$, there is a pinch-point singularity of the
dimer structure factor for $x$-oriented dimers, $S(\qq)$,
meaning that there is no divergence, but the limiting value at $\QQ$ depends 
on the direction of the ray along which it is approached.
The coefficient of this $k_y^2/(k_x^2+k_y^2)$ singularity is
$1/K$ according to Eq.~(\ref{eq:dimer-Sxx-dipolar}), so 
we can do a simple fit and call the result $K_S$.
Of course, the actual dependence on $\qq=\QQ+\kk$ must have
additions of higher order in $\kk$, since $S(\qq)$ is 
periodic in the Brillouin zone.  Therefore, only a small 
domain around $\QQ$ should be used in the fit, but it may 
be advantageous to use more than the wave-vectors immediately
adjacent to $\QQ$, as one can then extrapolate to $\QQ$
and eliminate most of the unwanted  additions. Of the four methods, 
this one is closest to direct measurement of the height Fourier 
spectrum $\langle |\tilde{h}(\kk)|^2\rangle$, which was the best 
method to extract stiffness constants from simulations of height 
models~\cite{Zeng,Raghavan} or random-tiling 
quasicrystals~\cite{Henley-RT,Oxborrow}.

In the RVB case, some additional steps are necessary, because we do 
not construct a height function and do not really even have a dimer 
configuration (recall that the contributions to the wave function from 
different dimer configurations are non-orthogonal and the simulations
sample pairs of dimer configurations). We only have the correlations 
$D_{xx}$ of an operator that has some projection onto a dimer-like 
variable as well as other contributions. This has two consequences for
$S(\qq)$. The first is that the ``other contributions'' contribute a 
constant background on top of the pinch-point singularity, which does 
not vanish even along the line $k_y=0$. That can in principle be remedied 
by fitting and subtracting off the constant addition, but unless the lattice
is very large such a procedure will not be perfect. In our fits carried 
out here, we simply use the value of the point that is next to the 
pinch point along line $k_y=0$ as our constant addition.

The second consequence of the lack of a formal height model
is that the measured $S(\qq)$ is a multiple of the assumed dimer 
structure factor by an unknown coefficient $c_S^2$. Fortunately, we can 
calibrate $c_S^2$ using the sectors with nonzero winding numbers, since 
the $\delta$-function peak at $\QQ$ in those cases (after subtracting 
the constant background) is proportional to $c_S^2$ times $(W_x^2+W_y^2)$ 
times known constants, allowing us to infer $c_S^2 \approx 0.56$. From this
value we can extract a normalized $S(\qq)$ and, finally, find the pinch-point 
coefficient we call $1/K_S$. This estimate of $K_S$ was computed for 
several system sizes and then extrapolated to $L=\infty$ by fitting functions 
$f(L)=a_0+a_1/L^2+a_2/L^3$ for the RVB and $f(L)=a_0+a_1/L^2$ for the CDM (i.e., 
with both forms not including the linear term). The results are given 
in Table~\ref{tab:grand}.

\subsection{Summary of the stiffness estimates}
\label{sec:sumK}

Table ~\ref{tab:grand} collects all four estimates of $K$,
with their statistical errors (one standard deviation). The fugacity 
$Z_2$ for long dimers specifies a family 
of RVB models and one of CDM models, with different exponents. 
Note that $K$ according to our convention is $\pi/8$ times $K$ as used
previously in Ref.~\onlinecite{Sandvik-Moessner}.

The respective estimates for the stiffness constant for a given case typically agree to within a few error bars.  
In some cases the deviations are larger than expected purely based on statistics. This is not unexpected, 
since the correlation functions we have analyzed are also affected by corrections to the leading forms we have used. 
Note that $K_S$ for the CDM with long dimers are systematically too large (the only really significant disagreement);
and $K_S$ for the RVB with long dimers appears to be slightly too large as well. Here the background contributions
which may not be perfectly subtracted off in our procedure, may be to blame.

The results for the CDM can be compared with the exact value $K_{\rm CDM} \equiv \pi/16 \approx 0.19635$, with which all $K$ 
estimates in Table ~\ref{tab:grand} agree to within 2 error bars or less. As another test, we calculated $K_P$ for the CDM with long bonds 
only (i.e., fugacities $Z_2=1$ and $Z_1=0$). The resulting value implies an exponent for the monomer correlations
of $\beta=0.11092(6)$, which agrees (within 1.5 error bars) with a previous obtained using a different analysis of
the monomer distribution function (and where it was conjectures that $\beta=1/9$).\cite{Sandvik-Moessner}

The good agreement between four different stiffness estimates provides strong evidence of an underlying height model 
description of the RVBs. The plausibility of the height-model approach for the RVB is partially motivated by the fact that 
the RVB and CDM coincide for SU($N$) spins when $N \to \infty$.\cite{Read2} 
One can then think of corrections to the continuum version of the height model 
for the CDM in terms of an $1/N$ expansion (which we have not carried out). 
The results discussed here show that the $1/N$ corrections all the way down to $N=2$ only correspond to a renormalization of the stiffness constant. 

\section{ORDER-PARAMETER DISTRIBUTION}
\label{vbshisto}

A columnar long-range ordered VBS on the square lattice breaks the translational and rotational lattice symmetries. As we have seen in the previous
sections, the RVB is a critical VBS with a rather slowly decaying dimer-dimer correlation function. This correlation function, Eq.~(\ref{dimer}), measures 
the magnitude of the VBS order parameter. In this section we look at another aspect of these critical VBS correlations, probing the individual order 
parameters for columns forming with $x$ and $y$ orientation of the modulated bonds, defined as
\begin{equation}
  \begin{aligned}
    D_x & = & \sum_{x=1}^{L}(-1)^x \sum_{y=1}^{L}  [ {\bf S}(x,y) \cdot {\bf S}(x+1,y) ]_{\rm conf},\\
    D_y & = & \sum_{y=1}^{L}(-1)^y  \sum_{x=1}^{L} [ {\bf S}(x,y) \cdot {\bf S}(x,y+1) ]_{\rm conf},
    \label{dxdy}
  \end{aligned}
\end{equation}
where $[...]_{\rm conf}$ indicates that these correlators are evaluated for an individual configuration (i.e., in the RVB they are matrix
elements between the sampled bra and ket states). The expectation values of these order parameters vanish. In the CDM, the dimer-dimer correlation 
functions that we investigated before correspond to their squares, i.e., the dominant structure factor in reciprocal space (as seen in Fig.~\ref{2dfft}) 
is $S(\pi,0)=\langle D_x^2\rangle/N$, and the behavior of this quantity as a function of the system size is shown in Fig.~\ref{diffw}. In the RVB, 
as we have discussed in Sec.~\ref{dimerdimer} and Fig.~\ref{diffw}, the squared order parameter based on the sampled values from Eq. (\ref{dxdy}) is not
exactly the same as the actual four-spin correlation function, but we have shown that the scaling properties are the same.

\begin{figure}
\centerline{\includegraphics[angle=0,width=8.4cm, clip]{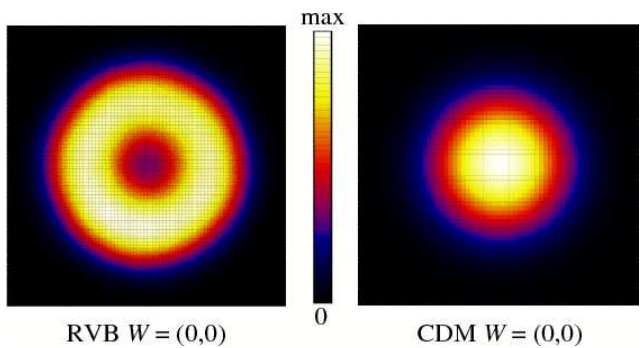}}
\caption{(Color online) VBS order parameter distribution function $P(D_x,D_y)$ in the space of point pairs $(D_x,D_y)$ generated in the Monte Carlo 
simulations of the RVB state (left) and CDM (right) for systems of size $L=64$. We are only concerned here with the shapes of these distributions 
(a ring with depleted weight in the center for the RVB and a broad central peak for the CDM) and have therefore not labeled the graphs with the range of 
$(D_x,D_y)$ or the actual values of the probability density.}
\label{vbshisto1}
\end{figure}

We here study the probability distribution $P(D_x,D_y)$ generated in the Monte Carlo sampling.
Each generated configuration of the valence bonds corresponds to pair of values $(D_x,D_y)$ evaluated according to the loop estimator Eq.~(\ref{scorr2}).
We use these to accumulate the histogram $P(D_x,D_y)$. Such histograms were generated by Sutherland in his loop-gas study,\cite{Sutherland} and
he noted a circular symmetry of the distribution (instead of a $4$-fold symmetry that would have been naively expected due to the lattice symmetry). 
At that time the results were affected by very large statistical uncertainties, however.

Dimer order-parameter histograms have recently become interesting in the context of {\it deconfined quantum critical} 
(DQC) points \cite{Senthil04,Sachdev08} in models exhibiting quantum phase transitions between the antiferromagnetic N\'eel state and a VBS state.\cite{Sandvik07,Lou09} 
A long-range ordered columnar  VBS corresponds to a distribution $P(D_x,D_y)$ peaked at one of the four points $(\pm |D|,\pm |D|)$, with the magnitude $|D|$ 
growing linearly with the system size $N=L^2$. In a finite system, in which the Z$_{4}$ symmetry is not broken, one expects equal weight in all these four 
peaks, as well as some weight between the peaks (which is related to the tunneling probability between the four ordered VBS states). As a DQC point is 
approached from the VBS side, one expects an emergent U($1$) symmetry in the system.\cite{Senthil04} This is manifested in $P(D_x,D_y)$ as a circular-symmetric 
distribution,\cite{Sandvik07,Sachdev08} i.e., for a finite system size $L$, the discrete four-fold (Z$_4$) symmetry naively expected for the VBS evolves 
into a continuous U($1$) symmetric distribution. For fixed couplings, the Z$_{\rm 4}$ symmetry develops as $L$ exceeds a length-scale characterizing
the spinon confinement (which diverges at the DQC point).

\begin{figure}
\centerline{\includegraphics[angle=0,width=8.4cm, clip]{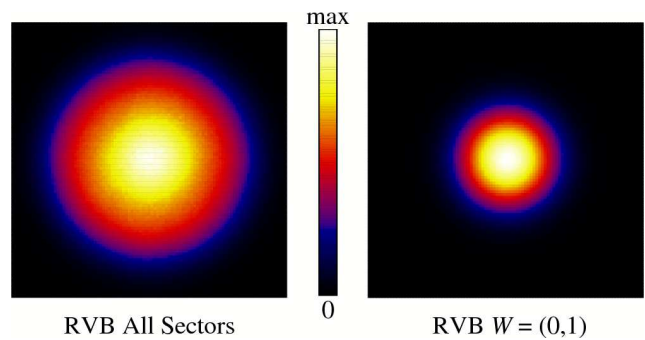}}
\caption{(Color online) VBS order parameter distribution function $P(D_x,D_y)$ for $L=48$ 
systems in the grand-canonical winding number ensemble (left) and with winding number $W=(0,1)$ (right).
Here the range of $D_x,D_y$ values is the same in both cases, i.e., the distribution for
$W=(0,1)$ is much narrower.}
\label{vbshisto2}
\end{figure}

While the RVB is a critical state, it does not correspond to a DQC point, because the spin correlations decay exponentially. At a DQC point, both the spin 
and dimer correlations are critical.\cite{Senthil04} It is nevertheless interesting to study the symmetry of the critical VBS order parameter in the RVB
and to compare it with the corresponding distribution in the CDM [where in ${\bf S}(x,y) \cdot {\bf S}(x+1,y)$ is replaced by the dimer occupation
number on the bond]. Results for $L=64$ systems in the winding number sector $W=(0,0)$ are shown in Fig.~\ref{vbshisto1}. Completely circular-symmetric 
distributions are seen in both cases, with no signs of $Z_{\rm 4}$ anisotropy. 
The natural expectation for a critical state 
is that the weight is centered around $(D_x,D_y)=(0,0)$, 
and this is in fact true for the CDM. 
Surprisingly, it is {\it not} true for the RVB critical state:
the distribution is instead ring shaped, with the dominant weight a finite radius away from the center. 
This is the behavior seen in candidate
models for DQC points in the VBS state close to the phase transition into the N\'eel state. 
The ring-shaped distribution in the RVB case is no contradiction to its being a critical state, because the ring's radius still 
grows slower with $L$ than $L^2$. The expectation value $\langle D^2\rangle/N$ is twice the structure factor $S(\pi,0)$ and hence 
grows as $L^{2-\alpha}$, with $\alpha \approx 1.20$ determined in Sec.~\ref{dimerdimer}.

In the case of a fixed non-zero winding number, the VBS order parameter is modulated by a plane wave, in the same way as its correlation
function is, as discussed in Sec.~\ref{dimerdimer}. Hence its spatial average
tends to cancel, with the result that the 
distribution function now has a central peak, 
as seen in Fig.~\ref{vbshisto2} (right panel) for $W=(0,1)$. 
For large winding numbers 
the distribution is marginally oval-shaped, reflecting the anisotropy induced by large
winding numbers (see Appendix \ref{app:height}). In Fig.~\ref{vbshisto2} 
the anisotropy is too small to observe clearly. 
Interestingly, when all winding numbers are included in grand-canonical simulations, the ring-shaped 
distribution seen for $W=(0,0)$ in Fig.~\ref{vbshisto1} no longer obtains. Although this sector completely dominates the grand-canonical ensemble 
(as seen in Fig.~\ref{Wprob}), the narrow central peaks contributed by the non-zero winding number sectors completely fill in the central portion 
inside the ring, resulting in a broad central peak, as shown in Fig.~\ref{vbshisto2} (left panel).

\section{Summary and Discussion}
\label{conclusion}

We have compared long-wavelength properties of short-bond RVB spin-liquid states with those of classical dimers,
specifically those associated with correlations and topological constraints of dimers. Taking properly into account the 
non-orthogonality of valence-bond basis states, arising from the internal bond-singlet spin structure which is not 
present in classical dimers, we have carried out numerically exact Monte Carlo simulations of the four-point correlation 
function measuring the tendency to formation of a VBS state. In contrast to the exponentially decaying two-point spin 
correlations,\cite{Liang} these VBS correlations decay as a power law. Such a power might have been anticipated based on the 
fact that the classical dimer-dimer correlations decay as $1/r^2$ (although the overcompleteness of the RVB could in
principle have led also to more dramatic deviations from the CDM), but the exact value of the exponent necessitates an 
exact treatment of the overcomplete basis, as we have done here. The result is that the correlations decay slower than 
what might have been anticipated, as $1/r^\alpha$ with $\alpha \approx 1.20$. 

The weighting of valence bond states is (qualitatively) different in that sampling the RVB state involves the 
transition graph of two states, whereas in the CDM only a single state is sampled (as different dimer configurations 
are by definition orthogonal). In particular, the loops are small in the short-bond RVB, as they necessarily must be 
in order to give exponentially decaying spin correlations (whereas in an  antiferromagnetically ordered 
state the typical loop size scales as the system size \cite{Beach2,Sandvik2}). The operators that we measure are 
also different in the two systems: the ``dimer-dimer'' correlations in the RVB actually refer to two-spin operators,
[Eq.~(\ref{bond})] in place of just bond occupation numbers in the CDM. We have confirmed that the changed $\alpha$ exponent 
(and presumably other changed expectations) in the RVB state originate solely from the different state weighting, not 
from the form of the correlation-function estimator Eq.~(\ref{scorr4}). 

The RVB structure factor has a ``pinch-point'' 
at $(\pi,\pi)$ in reciprocal space, in any winding number sector, 
like the well-known pinch-point in the CDM and other height models;
it further shows singularities related to the critical correlations
near to $(\pi,0)$ (but shifted by nonzero winding number) which are 
logarithmic for CDM at zero winding number, and otherwise are variable power 
laws. Finally, we found that introduced pairs of monomers, i.e., 
topological defects, are marginally (power-law) deconfined with a power law distribution 
of their separations.

Remarkably, all of the above observations fit into the framework
of the ``height model'' with a stiffness constant $K$
as worked out in Appendix ~\ref{app:height}.
Independent measurements of the stiffness constant can be
derived from 
(i) logarithms of the probabilities of
sectors with different winding numbers, 
(ii) the critical dimer correlation exponent, 
(iii) the monomer pair separation exponent, and
(iv) the pinch point of the structure factor $S({\qq})$.
All yielded $K_{\rm RVB} \approx 1.6 K_{\rm CDM}$.
Other behaviors, which do not yield measurements of $K$,
are also suggestive of this.
Thus, our results vindicate at last
the qualitative correctness of the zero-overlap 
assumption adopted in the RK QDM,
although quantitatively the RVB state has a larger
degree of VBS order (as expressed by that ratio of
stiffnesses $1.6$).
It is as if the RVB state were the ground state of the 
generalized RK state corresponding to some (still unknown) 
generalized classical dimer model.

We extended the model by introducing a small fraction of longer bonds 
(the next bipartite bond, which connects fourth-nearest neighbors). 
We studied the evolution of the power laws characterizing 
the dominant VBS correlations and monomer correlations as a function of 
the fugacity of long bonds. As in the CDM case, ~\cite{Sandvik-Moessner}, 
in the dimer-dimer correlations, a $(\pi,\pi)$ modulated ``dipolar'' term 
continues to have the $1/r^2$ behavior; on the other hand, 
a $(\pi,0)$ modulated ``critical'' term has an  increasing exponent, while
the monomer-monomer distribution function
has a decreasing exponent, both of which can be explained in terms
of a decreasing stiffness for the ``height'' fluctuations.
The monomers remain deconfined for all fugacities we studied.

We further studied the modifications to correlations
due to finite topological winding number, 
for both the RVB and classical dimers. 
The critical VBS correlations acquire a sinusoidal modulation,
correlations become anisotropic, and the effective stiffness is
increased, as expected from height-model calculations; 

We have also studied the joint probability distribution $P(D_x,D_y)$ of the VBS order parameters for columnar order with $x$ and $y$ oriented bonds. We found 
this distribution to be U($1$) symmetric, which in analogy with the proposed deconfined quantum-critical point \cite{Senthil04} should correspond to the 
lattice-imposed $Z_{\rm 4}$ symmetry of the VBS on the square lattice to be dangerously irrelevant [when regarded as a perturbation to an U($1$) symmetric 
field theory] in these critical systems (both in the RVB and the CDM). 
In a model that has one of these states as the ground state for some values of 
tunable parameters, e.g., the extended dimer models with ``Cantor deconfinement'' studied in Refs.~\onlinecite{Fradkin1}  and \onlinecite{Fradkin2}, 
one would then expect the U($1$) symmetry to be emergent upon approach to the critical point. 

Although we have here studied the RVB state without reference to any specific Hamiltonian, some general conclusions can still be drawn based on our results.
If a (local) Hamiltonian's ground state has algebraic correlations, then it must correspondingly have gapless excitations.  Thus, our results show that 
any Hamiltonian~\cite{Cano} with the RVB ground state is {\it gapless} in the singlet sector, even though it has a spin gap. Furthermore, the close 
qualitative correspondence of the RVB static correlations to the RK model \cite{RK1} suggests the long-wavelength excitations are similar too; 
these are known~\cite{Henley} to be coherent bosons with $q^2$ dispersion. Some actual spin systems may be spin gapped but singlet gapless. This has 
long been claimed for the spin-$1/2$ kagome antiferromagnet,\cite{Waldtmann,Jiang} although the spin gap is small enough that an extrapolated value 
of zero can not be ruled out.~\cite{Sindzingre} From this viewpoint, 
it is interesting to verify that the original
short-range RVB state has such a property. 

In experiments, the 2D organic $S=1/2$ spin-liquid candidate, 
EtMe$_3$Sb[Pd(dmit)$_2$]$_2$ shows gapless spin and singlet 
sectors in zero magnetic field,\cite{Matsuda-gapless}
but in a magnetic field, spin excitations become gapped while singlet excitations remain gapless and have high mobility, as indicated by specific heat 
and thermal conductivity. 

On the theory side, one might ask whether our result should have been expected.
Soon after the original proposal of the RVB wave function, 
field theorists argued that it corresponded to a $U(1)$ gauge theory,~\cite{Zheng-Sachdev,ioffe-larkin,Read2} and for a ``height model'' 
to be in its rough phase, as we found, is equivalent to being asymptotically a $U(1)$ gauge theory. But, the numerical value of the stiffness 
constant $K$ has not been measured previously (before our original estimate in Ref.~\onlinecite{yingabstract}); to our knowledge, it was not even suggested 
whether $K$ should be larger or smaller than $K_{CDM}$ of the QDM. If for no other reason, one must check the value of $K$ since, 
were it much larger, one would find long-range order in the dimer correlations (a spin-Peierls phase).

It would clearly be interesting to try to derive the height model (or the continuum version of it) starting from an $1/N$ expansion 
of the classical dimer model, which corresponds to the RVB for SU($N$) spins in the limit $N\to \infty$. Further, the recent construction 
\cite{Cano} of a model Hamiltonian which has exactly the RVB state studied here as its ground state also offers hope that one could actually, with 
extensions of that Hamiltonian, study a quantum phase transition in which the static properties of the critical point should be exactly those that we 
have investigated here in the RVB. 

\acknowledgments

We thank A. F. Albuquerque and F. Alet for communication 
related to pointing out an independent work that was carried out in 
parallel with ours.\cite{albu10} This work was 
supported by NSF Grants No.~DMR-0803510, No.~DMR-1104708 (AWS) and 
No.~DMR-1005466 (CLH). C.L.H. also acknowledges support from the 
Condensed Matter Theory Visitors Program at Boston University.

\appendix

\section{Four-spin correlators in the valence-bond basis}
\label{app:4spin-corr}

In this appendix we work out the loop expression for
four-spin correlators, analogous to the well-known 
two-spin expression Eq.~(\ref{scorr2}).

It is useful to consider the singlet projectors
\begin{equation}
C_{ij} = -({\bf S}_i \cdot {\bf S}_j-\hbox{$\frac{1}{4}$}).
\label{cij}
\end{equation}
When acting on a valence bond, this operator is diagonal with eigenvalue $1$. Denoting a singlet on sites 
$a$ and $b$ as $(a,b)$, we have
\begin{equation}
C_{ab}(a,b)=(a,b),
\label{cijdia}
\end{equation} 
whereas acting on a pair of different valence bonds leads to a simple reconfiguration of those bonds, e.g.,
\begin{eqnarray}
&&C_{bc}(a,b)(c,d)=\half (c,b)(a,d), \label{cijop1}\\
&&C_{bd}(a,b)(c,d)=\half (a,c)(b,d), \label{cijop2}
\end{eqnarray}
which can be shown easily by going back to the basis of $\up$ and $\dn$ spins.  Note the order of the indices within the singlets in Eq.~(\ref{cijop1}), which 
reflects consistently the chosen convention in the valence-bond state definition 
Eq.~(\ref{valpha}) when the sites $a,c$ are on sublattice $A$ and $b,d$ on 
sublattice $B$. We will also have to consider operations on two spins belonging to the same sublattice, as in Eq.~(\ref{cijop2}). We have not specified 
a convention for the order of the spins in singlets formed between two spins on the same sublattice, therefore, it is important to keep track of the signs, which
depends on the order in which the singlets are written. 

\begin{figure}
\center{\includegraphics[width=8cm]{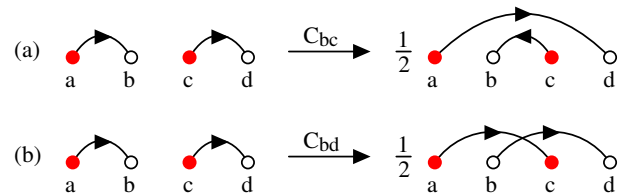}}
\caption{(Color online) Action of a singlet projection operator in two different cases; (a) when the sites $b$, $c$ are on different sublattices and (b) 
when $b$, $d$ belong to the same sublattice. The arrows indicate the order of the spins in a singlet; $(a,b)=(|\up_a\dn_b\rangle-|\dn_a\up_b\rangle)/\sqrt{2}$, 
and, in the case of spins on different sublattices, conforms with the definition 
Eq.~(\ref{valpha}) of bipartite valence bond states.}
\label{bondflip}
\end{figure}

\begin{figure}
\center{\includegraphics[width=8cm]{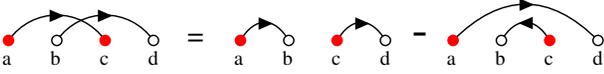}}
\caption{(Color online) Illustration of the equivalence Eq.~(\ref{vbequiv}), due to overcompleteness, between a state formed by two non-bipartite valence 
bonds and a superposition of two states involving only bipartite bonds.}
\label{vbconversion}
\end{figure}

Figure \ref{bondflip} illustrates the two different types of singlet projector outcomes in Eq.~(\ref{cijop1}) and Eq.~(\ref{cijop2}). 
In Fig.~\ref{bondflip}(a), both the initial 
and the final bond pairs are bipartite whereas in Fig.~\ref{bondflip}(b) 
the bonds after the operator has acted are non-bipartite. 
The non-bipartite bonds do not belong to 
the restricted basis of bipartite valence-bond basis in which we normally work. However, when generating non-bipartite bonds such as this (which can 
happen in the course of calculations), we can always rewrite them in terms of bipartite bonds. One can easily verify the following equivalence 
between valence bond pairs; 
\begin{equation}
(a,c)(b,d) = (a,b)(c,d)-(a,d)(c,b),
\label{vbequiv}
\end{equation}
which is illustrated in Fig.~\ref{vbconversion}. This relationship is particularly useful when sites $a,c \in A$ and $b,d \in B$, but it of course holds 
irrespective of sublattices.

As in Eq.~(\ref{szz}), we can take advantage of the spin-rotational symmetry also when considering a four-spin correlation function, writing the corresponding 
matrix element as
\begin{equation}
\langle V_\beta |({\bf S}_k \cdot {\bf S}_l)({\bf S}_i \cdot {\bf S}_j) |V_\alpha\rangle
=3\langle V_\beta |{S}^z_k{S}^z_l({\bf S}_i \cdot {\bf S}_j) |V_\alpha\rangle.
\label{cor4}
\end{equation}
Note, however, that we cannot further reduce this expression to a correlation function involving only $z$-spin components, because 
if $\gamma\not=z$, 
\begin{equation}
\langle V_\beta |S^z_kS^z_lS^z_iS^z_j|V_\alpha\rangle \not= \langle V_\beta |{S}^z_k{S}^z_l{S}^\gamma_i{S}^\gamma_j |V_\alpha\rangle.
\end{equation}
It is easy to see that the matrix element Eq.~(\ref{cor4}) is non-zero only if all four indices $i,j,k,l$ belong to the same loop, or if there are two 
indices in each of two loops. To carry out the calculations for these cases, it is convenient to make use of the singlet projection operator Eq.~(\ref{cij})
and write the matrix element as
\begin{eqnarray}
\label{szszcij}
&&\langle V_\beta |{S}^z_k{S}^z_l({\bf S}_i \cdot {\bf S}_j) |V_\alpha\rangle = \\
&&~~~~~~~~~~\hbox{$\frac{1}{4}$}\langle V_\beta |{S}^z_k{S}^z_l |V_\alpha\rangle -\langle V_\beta |{S}^z_k{S}^z_lC_{ij} |V_\alpha\rangle .\nonumber
\end{eqnarray}
We only go through the calculation in detail for the case where all four indices belong to the same loop, which is the most complicated situation. 

\begin{figure}
\center{\includegraphics[width=7cm]{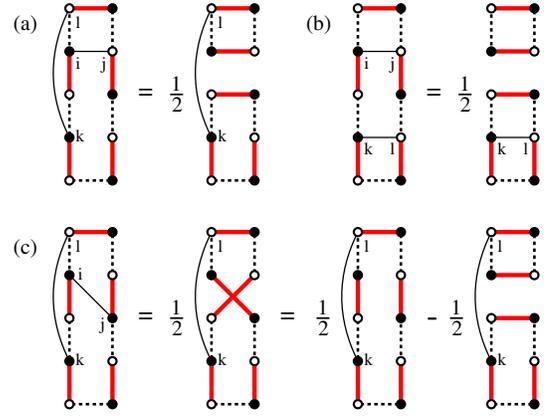}}
\caption{(Color online) Operations for evaluating the four-spin matrix element 
$\langle V_\beta | ({\bf S}_k \cdot {\bf S}_l)({\bf S}_i \cdot {\bf S}_j) | V_\alpha \rangle$ when all the sites $i,j,k,l$ are in the same loop of the transition 
graph. The thin lines connecting labeled sites refer to the operator components ${S}^z_k{S}^z_l$ and $C_{ij}$ in Eq.~(\ref{szszcij}). The solid and dashed bonds belong 
to $| V_\alpha \rangle$ and $\langle V_\beta |$, respectively.}
\label{sijkl}
\end{figure}

The procedure is illustrated in Fig.~\ref{sijkl}. Acting first with the singlet projector $C_{ij}$, the loop is split into two separate loops if $i,j$ are on different sublattices, 
as shown in Figs.~\ref{sijkl}(a) and \ref{sijkl}(b). If these sites are on the same sublattice, as in Fig.~\ref{sijkl}(c), the loop instead becomes ``twisted'' by two non-bipartite 
bonds. This loop can be re-cast in terms of two different contributions containing only bipartite bonds, by using the valence-bond equality illustrated in 
Fig.~\ref{vbconversion}. In each case, after $C_{ij}$ has acted, we can return to the spin representation of the valence bonds and evaluate the average of the 
remaining operator $S^z_kS^z_l$ exactly as we did for the two-spin correlation function. Here the result depends on whether $k,l$ are in the same loop (giving a 
non-zero correlation) or different loops (giving a zero average) after the loop-splitting with $C_{ij}$ has been enacted; these two different cases are illustrated 
in Fig.~\ref{sijkl}(a) and \ref{sijkl}(b) for the case $i,j$ in different sublattices [while for $i,j$ on the same sublattice, Fig.~\ref{sijkl}(c) only shows the case of $k,l$ 
in different parts of the split loop]. In  all cases, the matrix element ratio $\langle V_\beta |S^z_kS^z_lC_{ij}|V_\alpha\rangle/\langle V_\beta |V_\alpha\rangle$ 
is now easy to compute using Fig.~\ref{sijkl} and keeping in mind that an increased number of loops after a split by $C_{ij}$ increases the corresponding matrix 
element by a factor $2$ according to the loop expression Eq.~(\ref{overlap}) for the overlap. The four-spin correlation can then be extracted 
using Eqs.~(\ref{cor4}) and (\ref{szszcij}).

\begin{figure}
\center{\includegraphics[width=6.5cm]{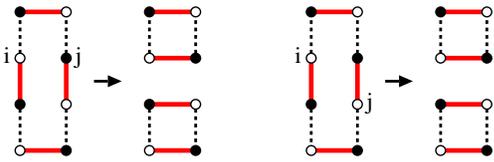}}
\caption{(Color online)
Subloops of a valence-bond loop with respect to two sites $i,j$. The ``cuts'' splitting the loop into subloops are at the solid bonds connected to $i$ and 
$j$ (which belong to the ket $|V_\alpha\rangle$; the state on which $C_{ij}$ acts), irrespective of the two possible locations of $i,j$ within these bonds. 
When $i,j$ are sites in the same bond in $|V_\alpha\rangle$, there is only a single subloop (the whole loop).}
\label{subloops}
\end{figure}

In order to write the final result in a compact unified form for all the different cases, it is useful to introduce the concept of {\it subloops} with respect to the 
operator $C_{ij}$ of a loop containing sites $i,j$, or $(i,j)$-subloops. As seen in Fig.~\ref{sijkl}, regardless of whether $i,j$ are on the same or different 
sublattices, the loop is split in the same way by $C_{ij}$ in all cases where such split loops appear. This can be formalized by the following convention: The 
splitting of a loop into $(i,j)$-subloops is accomplished using the bonds in the ket $|V_\alpha \rangle$ (the solid bonds in Fig.~\ref{sijkl}, on which $C_{ij}$ 
acts), i.e., the two $V_\alpha$-bonds on which $i,j$ are located are those that are reconfigured in such a way that the loop splits into two. The subloops then 
always contain only bipartite bonds. This definition is illustrated in Fig.~\ref{subloops}. We also introduce a symbol to distinguish between the cases of 
$k,l$ in the same subloop or different subloops;
\begin{equation}
\delta^{kl}_{ij} = \left \lbrace \begin{array}{l}
0,~~\hbox{for $k,l$ in the same $(i,j)$-subloop}, \\
1,~~\hbox{for $k,l$ in different $(i,j)$-subloops}.
\end{array}\right. 
\label{deltaijkl}
\end{equation}
If $i,j$ are on the same bond of $|V_\alpha \rangle$, $C_{ij}$ does not change the loop and there is then only a single subloop (the intact original loop) 
and $\delta^{kl}_{ij}=0$ for all $k,l$.

The remaining cases of non-zero four-spin matrix elements involve two loops (with two indices in each loop). These calculations are easier than the case of all 
indices in the same loop, because there are no subloops to consider, and we just list the results. The full final result for all non-zero four-spin matrix 
elements is given in the main text as Eq.~(\ref{scorr4}).

Note that whereas the sign of the two-spin correlation Eq.~(\ref{scorr2}) is always dictated by the staggered phase factor, 
the sign of the four-spin correlation is different from the four-site staggered phase $\phi_{ij}\phi_{kl}$ if all 
the indices are in the same loop and $k,l$ belong to different $(i,j)$-subloops.

The concept of subloops may seem unnecessarily complicated in the definition of $\delta^{ij}_{kl}$ in Eq.~(\ref{deltaijkl}), since this number ($0$ or $1$) is 
essentially also determined by the order in which the indices $i,j,k,l$ appear when traversing a loop. If only one of the indices $k,l$ appear between $i,j$, 
then, in most cases, $k,l$ are in different subloops and $\delta^{ij}_{kl}=1$. There are, however, special cases where the definition based on the order of 
indices is ambiguous, e.g., when they are all on the same valence bond in the ket $|V_\alpha\rangle$. In that case, $k,l$ are in the same subloop and
$\delta^{ij}_{kl}=0$, as also explained in Fig.~\ref{subloops}.

\section{Calculations based on height representation}
\label{app:height}

Any {\it complete} covering of a bipartite planar lattice
(such as the square lattice) by dimers can be mapped into
a configuration of ``heights'' representing a kind of
interface model.  Often , the ensemble weighting
corresponds to the ``rough'' phase of the interface.
In this case, many statistical properties may be 
derived from a simple (Gaussian) classical field theory
in terms of the coarse-grained height function, using
the ``Coulomb-gas'' formalisms introduced in the 
Kosterlitz-Thouless theory of the two-dimensional 
XY model.\cite{Nelson,Nienhuis}  Bipartite dimer
coverings are a subset of a larger class of ``height''
models treated by this formalism, which also include
random-tiling quasicrystals.\cite{Henley-RT,Oxborrow}

The CDM is known to be in this ``rough'' phase.
In the case of the RVB wave function, for which this property had 
not been known, it is shown in this paper that all statistical behaviors 
are consistent with a rough height model.  
It should be emphasized that this is an emergent behavior,
since there is no exact way to map spin states to dimer
coverings (the dimers to spins mapping is not invertible).
We might hypothesize the existence of some hidden, nonlocal
way to define winding numbers and perhaps height fields
from the spins;  however, the nonzero overlap between configurations
in different winding-number sectors (see Fig.~\ref{Wprob})
shows that there can not be an exact mapping of that sort.

The starting point of the height treatment is that
the probability of a (coarse-grained) height field
$\{ \hh(\rr) \}$ is given by $\exp[-\Ftot(\{\hh(\rr)\})]$,
where 
\beq
    \Ftot = \int d^2 \rr \frac{1}{2} K |\nabla \hh(\rr)|^2.
    \label{eq:Ftot-h}
\eeq
We here study various consequences following from this.    

\subsection{Relation of height field and dimer operators}

There are two closely related ways to define a height function,
for a dimer model, as laid out in Ref.~\onlinecite{Zeng}.
The microscopic height $\hraw(\rr)$ is 
defined on dual vertices (centers of plaquettes);
we set 
   \begin{subequations}
       \label{eq:hraw}
   \begin{eqnarray}
       \hraw(x+\half,y+\half)&-&\hraw(x-\half,y+\half)\nonumber\\ 
               &=&(-1)^{x+y}[4 n_y(x,y)-1],\\*
       \hraw(x+\half,y+\half)&-&\hraw(x+\half,y-\half)\nonumber\\
               &=&(-1)^{x+y}[4 n_x(x,y)-1].
   \end{eqnarray}
   \end{subequations}
Thus $\hraw$ takes a step $\pm 3$ across a dimer, or $\mp 1$ across an 
unoccupied bond, where the sign alternates between even and
odd vertices of the lattice.  If one takes four steps around a
vertex, one crosses a dimer once and an unoccupied bond 
three times such that the net difference is zero, ensuring
a well-defined height field.

A second, locally averaged height function $\hav(x,y)$
is defined on the original vertices, being  the mean of
$\hraw$ on the four surrounding plaquettes. 
[Note the locally averaged $\hav(x,y)$ is
not quite identical to the fully coarse-grained 
height function assumed in the field theory, although
we use the same notation $\hh(\rr)$.]
This $\hav(x,y)$ is uniform in any one of the four 
special domains in which the dimers are aligned 
on opposite sites of plaquettes; it shifts by one
unit on crossing a domain wall to the next domain.
A change of $\pm 4$ in $\hav$ brings us back to the
same domain.

Thus, the dimer occupation can be written as a period-four
function of the local height variable, 
   \begin{subequations}
   \label{eq:ndimer-hh}
   \begin{eqnarray}
      n_x(\rr) &=& \half \big[ \cos\big(\frac{2\pi \hh}{4}\big)^2  +
                       (-1)^x \cos\big(\frac{2\pi \hav}{4}\big)\big],~~~~\\
      n_y(\rr) &=& \half \big[ \sin\big(\frac{2\pi \hh}{4}\big)^2 +
                       (-1)^y \sin\big(\frac{2\pi \hav}{4}\big)\big].~~~~
   \end{eqnarray}
   \end{subequations}
The configurations with a given winding number may be visualized as
fluctuating domains with smoothed domain walls. 
For winding number $W=(W_x,0)$, a net number of domain walls $4W_x$ 
must be crossed as the system is traversed in the $x$ direction.
There is no long-range dimer order,
so the domain walls thereby enforced are delocalized; indeed, in a snapshot of the
configuration, they are lost in the dense array of random domain walls which are
part of the inherent fluctuations even in the $W=(0,0)$ sector.

\subsection{Effects of long dimers}

In the present simulations, sometimes dimers are 
permitted (both in CDM and RVB models) between sites 
separated by a $(2,1)$ type vector with a fugacity $Z_2$.
This requires us to modify the height construction.
Say this dimer extends from (0,0) to (2,1).  The height
changes across the lattice edges (0,0)--(1,0) and (1,1)--(2,1) as
if there were ordinary dimers occupying both edges 
(i.e. $-1$ times the height change if those edges were vacant.)  
As for the lattice edge (1,0)--(1,1) bisected by the long dimer,
the height change is $+5$ times the height change the
vacant edge would have had.  Around the vertex (1,0) or (1,1),
the net height changes are $3+3-5-1=0$, showing the modified 
construction is well defined.

It can be seen that long dimers allow larger differences in
height between adjacent sites.  In the coarse-grained picture
this means that height gradients are penalized less and
thus $K$ is decreased.  Indeed, it was observed in
previous work~\cite{Sandvik-Moessner} that in the CDM 
when {\it only} long dimers are present, $K$ is reduced
by a factor of $2/9$.

\subsection{Dimer correlations: dipolar term}
\label{sec:app:dimer-dipolar}

It seems as if Eqs.~(\ref{eq:hraw}) and (\ref{eq:ndimer-hh})
express contradictory relations between the height field and the dimer
configuration. The proper resolution is that the dimer field has
two slowly varying parts that are modulated in different ways
with respect to the lattice, 
   \begin{subequations}
   \label{eq:dimer-h-grand}
   \begin{eqnarray}
    &&  n_x(x,y)-\quarter \approx (-1)^{x+y} \frac{d\hh}{dy}  \\
          &&~~~~~~~~~~~~~~~~~~~~ + \frac{(-1)^x}{2} \cos\Big(\frac{2\pi \hav}{4}\Big),~~~~~ \nonumber \\
    &&  n_y(x,y)-\quarter \approx (-1)^{x+y+1} \frac{d\hh}{dy}  \\
          &&~~~~~~~~~~~~~~~~~~~~  + \frac{(-1)^y}{2} \sin\Big(\frac{2\pi \hav}{4}\Big),~~~~~  \nonumber
   \end{eqnarray}
   \end{subequations}
which is equivalent to Eq.~(2.4) of Ref.~\onlinecite{Fradkin1}.
It turns out that the $n_x$--$n_x$ 
dimer occupation correlation, as a function of displacement $\rr=(x,y)$,
breaks up into two slowly decaying terms, 
$D_{xx}(\rr) = D_{xx}^{\rm dip}(\rr) + \Dxxcrit (\rr)$, 
which are due to the two kinds of terms in Eqs.~(\ref{eq:dimer-h-grand}).

Consider the first kind of term.
Equation (\ref{eq:Ftot-h}) implies, for the Fourier transform of the
height field, $\langle | \tilde{h}(\qq)|^2 \rangle \approx 1/K|\qq|^2$
for small wavevectors $\qq$.  Combining with the
$\hh$ gradient terms in Eq.~(\ref{eq:dimer-h-grand}),
we find 
   \beq
      S(\QQ+\kk) \approx \frac{k_y^2}{K|\kk|^2}
   \label{eq:dimer-Sxx-dipolar}
   \eeq
for the $x$-dimer structure factor near $\QQ=(\pi,\pi)$.
Taking the Fourier transform of Eq.~(\ref{eq:dimer-Sxx-dipolar})
gives the (two-dimensional) pseudo-dipolar correlations 
   \beq
      D_{xx}^{\rm dip} (\rr) \approx 
   (-1)^{x+y} {\rm Const} \frac{x^2-y^2}{2\pi K |\rr|^4}.
   \eeq
The radial dependence of this is $1/r^2$ in any direction, irrespective
of the value of $K$.

\subsection{Dimer correlations: Critical term}
\label{sec:app:dimer-critical}

We now turn to the second kind of term in Eqs.~(\ref{eq:dimer-h-grand}),
the terms periodic in $\hav$.
By a calculation standard in height-model literature~\cite{Raghavan,Zeng},
they imply the Coulomb gas (critical) term, 
   \beq
     \Dxxcrit (\rr) \propto  \frac{(-1)^x }{|\rr|^\alpha},
   \label{eq:C-crit-decay}
   \eeq
where 
   \beq
         \alpha = \frac{(2 \pi /4)^2}{2\pi K} \equiv \frac{\pi}{8K}.
         \label{eq:alpha}
   \eeq
It is a peculiarity of the CDM, with nearest-neighbor dimers and
equally weighted configurations, that $\alpha=2$.
Thus {\it both} terms have the same decay exponent 
and in fact they cancel exactly on certain sites.
Modifying the relative weighting of dimer configurations normally
changes $\alpha$.  If $\alpha < 1/4$, the height configuration 
locks into a flat state (roughening transition) which means that 
the dimers lock into a long-range ordered state.  However, 
in this study, $\alpha$ is reduced from the CDM 
value of 2 by a relatively modest amount.

The same kind of calculation implies that 
  \beq
    \Dxxcrit (L/2,L/2) \propto \frac{1}{L^\alpha},
  \eeq
with the same $\alpha$ as in Eq.~(\ref{eq:C-crit-decay}), but a different
prefactor.  Note that (so long as the elasticity is isotropic)
the dipolar contribution $D_{xx}^{\rm dip} (\rr)$ is 
exactly zero along the lines $x=\pm y$ (even as its
asymptotic $r$ dependence breaks down) and therefore does not
contribute to $D_{xx}(L/2,L/2)$.

\subsection{Topological (monomer) defects and their correlations}
\label{sec:app-monomers}

If a site is uncovered, the height differences do {\it not}
cancel in going around it, but change by $b=\pm 4$ (where the
sign depends on whether the vertex is even or odd).
Such defects can only be created in pairs of opposite
charge, and play the same role  as vortices in the 
Kosterlitz-Thouless theory.  The $K$ values in our
simulations are small enough that we are above the
Kosterlitz-Thouless unbinding transition, i.e., 
if there were nonzero fugacity to have defects, they would destroy
the critical state at sufficiently long length scales.
However, the fugacity is in fact zero (except that in
some simulations, one pair is inserted by hand as a probe).

The presence of a defect at (say) the origin
enforces a background gradient in the height
field with $|\nabla\hh| = b/2\pi r$.
When substituted into Eq.~(\ref{eq:Ftot-h}), 
that would give a logarithmically divergent total,
except that the divergence gets cut off by another
defect at distance $R$.  The result is that
the effective potential cost for the defects to
be separated by $R$ is $(K/2\pi) b^2 \ln R$, and
the pair distribution is given by
  \beq
        M(R) \propto \frac{1}{R^\beta},
  \eeq
with
   \beq
      \beta  = \frac{K b^2}{2\pi} = \frac{8 K}{\pi},
     \label{eq:beta}
  \eeq
and in particular $\beta=1/2$ for the basic CDM.

\subsection{Sector probabilities}
\label{sec:app-Wprob}

We now turn to the effects of enforcing net winding numbers
$W_x,W_y$. This is equivalent to a boundary condition that
$\hh(L,y)\equiv \hh(0,y)+ 4 W_x$ and $\hh(x,L)\equiv \hh(x,0)+ 4 W_y$.
In light of Eq.~(\ref{eq:ndimer-hh}), no discontinuity is implied
in the actual dimer pattern, since that depends on $\hh(\rr)$
with period 4.  It would be exactly analogous to enforcing,
in an $XY$ model, angle differences $(2\pi W_x, 2\pi W_y)$
across the system.

Thus the effect of winding number $(W_x,W_y)$ is to 
impose a uniform ``background'' height tilt 
$(m_x,m_y) = 4 (W_x,W_y)/L$.  We write 
   \beq
      \hh(\rr) =  m_x x + m_y y + \hh'(\rr),
      \label{eq:h-background})
   \eeq
separating the height field into the background plus a (smaller) deviation $\hh'(\rr)$ that 
satisfies periodic boundary conditions.  

If we substitute the free energy Eq.~(\ref{eq:Ftot-h}) into 
Eq.~(\ref{eq:h-background}), we see that 
   \beq 
           \Ftot(\{\hh\}) =  \Ftot (\{\hh'\}) + \Delta F(W_x,W_y),
   \label{eq:Ftot-hprime}
   \eeq
where 
   \beq
            \Delta F(W_x,W_y)  = \half K L^2 (m_x^2+m_y^2) = 8 K (W_x^2+W_y^2).
   \eeq
Since $\Ftot$ in Eq.~(\ref{eq:Ftot-hprime}) is exactly the same function as before, it
follows that when we integrate over all configurations of $\{\hh'(\rr)\}$ to obtain
the partial partition function $Z(W_x,W_y)$ for a given sector, 
$Z(W_x,W_y)= Z(0,0) \exp[-\Delta F(W_x,W_y)]$.  We conclude that the relative
probabilities of different sectors are given by
   \beq
             P(W_x,W_y) = P(0,0) e^{- 8 K(W_x^2+W_y^2)}.
   \label{eq:P-W}
   \eeq
In checking the normalization of $P(W_x,W_y)$, it should be remembered that
e.g. the (1,0) sector is fourfold degenerate [the possible winding numbers
are $(\pm1,0)$ and $(0,\pm 1)$], as are the (1,1) and (2,0) sectors.

\subsection{Correlation modulation due to winding number}
\label{app:winding-corr}

To calculate the critical contribution in the presence of a 
background $\hh$ gradient associated with a winding number, 
we merely need to substitute Eq.~(\ref{eq:h-background}) into
Eqs.~(\ref{eq:dimer-h-grand}), remembering that the 
rightmost terms are the ones contributing to the desired
correlation.  The result is that we get the correlation
due to the $\hh'$ field (i.e. the same as before) times
$\cos[\frac{2\pi}{4}(m_x x + m_y y)]$, where $(x,y)$ is
the vector connecting the two points.  In other words,
   \beq
           \Dxxcrit(\rr;W)= \Dxxcrit(\rr;0) \cos(\delta \QQ \cdot \rr),
   \label{eq:Dxxcrit-modulated}
   \eeq
where $\Dxxcrit(\rr;W)$ means $\Dxxcrit(\rr)$ given winding numbers $W$,
and
   \beq
             \delta \QQ \equiv \frac{2\pi}{4}(m_x,m_y) = {2\pi} (W_x,W_y)/L.
   \label{eq:delta-Q}
   \eeq
Since $\Dxxcrit(\rr;0)$ already includes a $(-1)^x$ modulation, it
follows that the structure factor singularity of $\Dxxcrit(\rr;W)$
gets shifted to 
   \beq
          \QQ= (\pi,0)\pm \delta \QQ.
   \eeq

\subsection{Anisotropic effects due to winding number}
\label{app:aniso}

In a height model, the free-energy density
is a function of $\nabla \hh(\rr)$ and its derivatives, 
satisfying all lattice symmetries.  
The free-energy density in Eq.~(\ref{eq:Ftot-h}) is the
lowest term of its Taylor expansion in $\nabla \hh$.
The next terms consistent with the square lattice are quartic, thus,
the free-nergy density becomes
   \begin{eqnarray}
     f(\nabla \hh) = \half K |\nabla\hh|^2 &+&
g_{11} \Bigg[ \Big(\frac{d\hh}{dx}\Big)^4 +\Big(\frac{d\hh}{dy}\Big)^4 \Bigg],
  \nonumber \\
&+& 2 g_{12} \Big(\frac{d\hh}{dx}\Big)^2 \Big(\frac{d\hh}{dy}\Big)^2 .
   \label{eq:f-quartic}
   \end{eqnarray}

If we insert Eq.~(\ref{eq:f-quartic}) into Eq.~(\ref{eq:h-background})
The effective free energy density to 
lowest order in $\hh'$ is
   \beq
     f = \frac{1}{2} K_x \Big(\frac{d\hh}{dx}\Big)^2 + \frac{1}{2} K_y \Big(\frac{d\hh}{dy}\Big)^2 
      + K_{xy} \Big(\frac{d\hh}{dx}\Big)\Big(\frac{d\hh}{dy}\Big), 
   \eeq
where
   \begin{subequations}
   \label{eq:K-gij}
   \begin{eqnarray}
           K_x &\equiv& K +  12 g_{11} m_x^2 + 2 g_{12} m_y^2, \\
           K_y &\equiv& K +  12 g_{11} m_y^2 + 2 g_{12} m_x^2, \\
           K_{xy} &\equiv&   4 g_{12} m_x m_y.
   \end{eqnarray}
   \end{subequations}
The nonlinear terms of a background tilt were
considered and measured from simulations in
the quasicrystal random tiling context~\cite{Oxborrow}.
It is possible, in principle, to extract analytical expressions for 
the nonlinear terms from the exact solutions.

Next we consider how this modifies correlations.
For simplicity, consider the case $m_y=0$.
We make a change of variables 
   \beq
      x'\equiv \gamma x;  \quad y' \equiv \gamma^{-1} y,
   \eeq
where
    \beq
        \gamma \equiv (K_x/K_y)^{1/4}.
    \eeq
In the new coordinates, the  free energy density is
    \beq
    f = \frac{1}{2} K'  
\Bigg[ \Big(\frac{d\hh'}{dx'}\Big)^2 + \Big(\frac{d\hh'}{dy'}\Big)^2 \Bigg],
    \label{eq:f-transformed}
    \eeq
with an effective stiffness $K' \equiv \sqrt{K_x K_y}.$
In these new coordinates, Eq.~(\ref{eq:f-transformed}) looks isotropic again
and the same results must follow for the behavior of all correlations.
In particular, the dimer and monomer correlation decay exponents,
$\alpha$ and $\beta$, depend on $K'$ in the same way they previously 
did on $K$.
In the general case that $m_x m_y \neq 0$, the effective stiffness is
   \beq
            K' \equiv \sqrt{K_x K_y - K_{xy}^2}.
   \label{eq:Keff}
   \eeq
For small $W/L$, i.e. small $(m_x,m_y)$, this reduces
in light of Eqs.~(\ref{eq:K-gij}) to 
to $K'\approx K+ 96 (g_{11}+g_{12})(W_x^2+W_y^2)/L^2$.
Hence large $L$,  and a winding number $W$
the corrections to exponents scale the same way, $\delta \alpha \sim \delta \beta \sim W^2/L^2$.

Notice that the decay exponent is the same in all spatial directions.  
The way the anisotropy gets expressed in the correlations with 
variable exponents is that (e.g.) dimer correlations 
do not fall off exactly as $1/r^\alpha$ , but rather as $1/{r'}^{\alpha}$, 
where $r'\equiv \sqrt{\gamma^2 x^2 + \gamma^{-2} y^2}$, and
similarly for monomer pair separations.
It would be interesting to see whether the anisotropy of {\it spin} correlations,
as shown in Fig.~\ref{spincorr}, is expressed by the same ratio $\gamma$.


\begin{thebibliography}{00}
\bibitem{monomernote}
In the CDM, the monomer distribution is directly given by the "end-end" separation of the loop under construction (i.e., the distance between the 
two defects introduced).\cite{Adams,Sandvik-Moessner} In the RVB, the defects 
generated when making the loops have spin, and the statistics of 
these spinful defects is not simply related to that of monomers.
\end{thebibliography}
\end{document}